\documentclass[11pt,letterpaper]{article}

\usepackage[dvipsnames,table]{xcolor}

\usepackage{latexsym,multirow}
\usepackage{amssymb,amsmath, bm, centernot}
\usepackage{graphicx}
\usepackage{verbatim}
\usepackage{centernot}
\usepackage{multicol}
\usepackage{mathrsfs}
\usepackage[export]{adjustbox}
\usepackage{centernot}
\usepackage{caption}
\usepackage{enumitem}
\usepackage{pgf}
\usepackage{tikz}
\usetikzlibrary{positioning,backgrounds,snakes}
\usepackage{pgfplots}
\usepackage{footmisc}

\usepackage{lscape}

\usepackage[backend = biber, natbib=true, style=authoryear, uniquelist=false,maxcitenames=1]{biblatex}
\usepackage[pdftex, bookmarksopen=true, bookmarksnumbered=true,
pdfstartview=FitH, breaklinks=true, urlbordercolor={0 1 0},
citebordercolor={0 0 1},  hidelinks]{hyperref}
\usepackage{colortbl}
\usepackage{subfigure}
\usepackage{inputenc}
\usepackage{dcolumn}
\newcolumntype{.}{D{.}{.}{-1}}
\newcolumntype{d}[1]{D{.}{.}{#1}}
\usepackage{ntheorem}
\makeatletter
\renewtheoremstyle{plain} % Adds automatic line break, if heading is too long
  {\item{\theorem@headerfont ##1\ ##2\theorem@separator}~}
  {\item{\theorem@headerfont ##1\ ##2\ (##3)\theorem@separator}~}
\makeatother

\theorembodyfont{\normalfont}
\theoremheaderfont{\normalfont\bfseries}

\theoremstyle{break}
\newtheorem{assumption}{Assumption}
\theoremstyle{definition}
\theoremstyle{break}
\newtheorem{definition}{Definition}

\theoremstyle{definition}
\theoremstyle{break}
\newtheorem{setting}{Setting}

\theoremstyle{break}
\theoremstyle{plain}

\theoremstyle{break}
\newtheorem{theorem}{Theorem}

\theoremstyle{break}
\theoremstyle{plain}
\newtheorem{result}{Result}
\newtheorem{remark}{Remark}
\newcommand{\qed}{\hfill \ensuremath{\Box}}
\newcommand{\indep}{\mbox{$\perp\!\!\!\perp$}}

\usepackage{rotating}

\usepackage[compact]{titlesec}
\usepackage{array}
\newcolumntype{L}[1]{>{\raggedright\let\newline\\\arraybackslash\hspace{0pt}}m{#1}}
\newcolumntype{C}[1]{>{\centering\let\newline\\\arraybackslash\hspace{0pt}}m{#1}}
\newcolumntype{R}[1]{>{\raggedleft\let\newline\\\arraybackslash\hspace{0pt}}m{#1}}

\usepackage{arydshln}

\usepackage{booktabs}

\usepackage{tcolorbox}

\usepackage{framed}

\allowdisplaybreaks

\begin{document}

% === new commands ===
\newcommand\ud{\mathrm{d}}
\newcommand\dist{\buildrel\rm d\over\sim}
\newcommand\ind{\stackrel{\rm indep.}{\sim}}
\newcommand\iid{\stackrel{\rm i.i.d.}{\sim}}
\newcommand\logit{{\rm logit}}
\renewcommand\r{\right}
\renewcommand\l{\left}
\newcommand\pre{{(t-1)}}
\newcommand\cur{{(t)}}
\newcommand\cA{\mathcal{A}}
\newcommand\cB{\mathcal{B}}
\newcommand\bone{\mathbf{1}}
\newcommand\E{\mathbb{E}}
\newcommand\Var{{\rm Var}}
\newcommand\cD{\mathcal{D}}
\newcommand\cK{\mathcal{K}}
\newcommand\cP{\mathcal{P}}
\newcommand\cT{\mathcal{T}}
\newcommand\cX{\mathcal{X}}
\newcommand\cXR{\mathcal{X,R}}
\newcommand\wX{\widetilde{X}}
\newcommand\wT{\widetilde{T}}
\newcommand\wY{\widetilde{Y}}
\newcommand\wZ{\widetilde{Z}}
\newcommand\bX{\mathbf{X}}
\newcommand\bx{\mathbf{x}}
\newcommand\bT{\mathbf{T}}
\newcommand\bt{\mathbf{t}}
\newcommand\bwT{\widetilde{\mathbf{T}}}
\newcommand\bwt{\tilde{\mathbf{t}}}
\newcommand\bbT{\overline{\mathbf{T}}}
\newcommand\bbt{\overline{\mathbf{t}}}
\newcommand\ubT{\underline{\mathbf{T}}}
\newcommand\ubt{\underline{\mathbf{t}}}
\newcommand\bhT{\widehat{\mathbf{T}}}
\newcommand\bht{\hat{\mathbf{t}}}

%% == Network Terminology ==
\newcommand\cF{\mathcal{F}} %% Friends network
\newcommand\cC{\mathcal{C}} %% Classmate network
\newcommand\cS{\mathcal{S}} %% sample space
\newcommand\cN{\mathcal{N}} %% Neighbors
\newcommand\bZ{\mathbf{Z}} %% Treatment
\newcommand\bz{\mathbf{z}} %% Realized Treatment

\newcommand\bW{\mathbf{W}} %% Treatment
\newcommand\bY{\mathbf{Y}} %% Treatment

\newcommand\bC{\mathbf{C}} %% Treatment
\newcommand\bc{\mathbf{c}} %% Treatment

\newcommand\bV{\mathbf{V}} %%
\newcommand\bv{\mathbf{v}} %%
\newcommand\bbv{\mathbf{\bar{v}}} %%
\newcommand\bbx{\mathbf{\bar{x}}} %%

\newcommand\bd{\mathbf{d}} %%
\newcommand\bP{\mathbf{P}} %%
\newcommand\bu{\mathbf{u}} %%

\newcommand\cw{\mathcal{w}} %%
\newcommand\cW{\mathcal{W}} %%

\newcommand\bg{\bar{g}} %% bar g
\newcommand\pg{g^\prime} %% bar g

\newcommand\bw{\mathbf{w}} %% bar g

\newcommand\cG{\mathcal{G}} %%  network 1
\newcommand\cH{\mathcal{H}} %% network 2
\newcommand\cU{\mathcal{U}} %% network 3

\newcommand\pS{\mathscr{S}}
\newcommand\pP{\mathscr{P}}

\newcommand\cM{\mathcal{M}} %% a set of networks
\newcommand\cO{\mathcal{O}} %%  a set of networks
\newcommand\cJ{\mathcal{J}} %%  a set of networks

\newcommand\CDE{{\rm {\bf CDE}}}
\newcommand\ADE{{\rm {\bf ADE}}}
\newcommand\cADE{{\rm {\bf cADE}}}
\newcommand\DE{{\rm {\bf DE}}}
\newcommand\cANSE{{\rm {\bf cANSE}}}
\newcommand\ANSE{{\rm {\bf ANSE}}}
\newcommand\ASE{{\rm {\bf ASE}}}
\newcommand\MSE{{\rm {\bf MSE}}}
\newcommand\INT{{\rm {\bf INT}}}
\newcommand\ATSE{{\rm {\bf ATSE}}}
\newcommand\MDE{{\rm {\bf MDE}}}
\newcommand\CSE{{\rm {\bf CSE}}}
\newcommand\TSE{{\rm {\bf TSE}}}

\newcommand\Prz{{\rm Pr}_{\mathbf{z}}}

\newcommand\MR{{\rm MR}}
\newcommand\RR{{\rm RR}}

\newcommand\Nor{{\rm Normal}}

%% == Naoki included this ===
\newcommand\cATIE{{\sc cATIE}}
\newcommand\ATIE{{\sc ATIE}}
\newcommand\cg{\cellcolor[gray]{0.7}}

\newcommand\mo{\mathbf{1}}
\newcommand\PA{\mbox{PA}}
\newcommand\PAo{\mbox{PA}^\ast}
\newcommand\sign{\texttt{sign}}

\newcommand{\argmax}{\operatornamewithlimits{argmax}}
\newcommand{\argmin}{\operatornamewithlimits{argmin}}

\newcommand{\minb}{\operatornamewithlimits{min}}
\newcommand{\maxb}{\operatornamewithlimits{max}}

\newcommand\spacingset[1]{\renewcommand{\baselinestretch}%
  {#1}\small\normalsize}

\newcommand{\nec}[1]{\textbf{\textcolor{magenta}{(NE: #1)}}}

\spacingset{1.25}

\newcommand{\tit}{
  \mbox{ {Covariate Selection for Generalizing Experimental
      Results:}} \\
  Application to Large-Scale Development Program in Uganda}

%%%%%%%%%%%%%%%%%%%%%%%%%%%%%%%%%%%%%%%%%%%%%%%%%%%%%%%%%%%%%%%%%%%%%%%%

% \if0\blind

\title{\tit \thanks{We would like to thank Christopher Blattman,
    Nathan Fiala, and Sebastian Martinez for sharing their data. We
    are also grateful to Alexander Coppock, Don Green,
    Chad Hazlett, Zhichao Jiang, and Soichiro Yamauchi as well as the
    participants of the MPSA, the UCLA IDSS workshop and the Yale Quantitative
    Methods Workshop, for their helpful comments on an earlier version
    of the paper.  We also thank the editor and two anonymous
    reviewers for providing us with valuable comments. \vspace{0.05in}}}

\spacingset{1.0}
\author{Naoki Egami\thanks{Assistant Professor, Department of
    Political Science, Columbia University, New York NY  10027. Email:
    \href{mailto:naoki.egami@columbia.edu}{naoki.egami@columbia.edu},
    URL: \url{https://naokiegami.com}\vspace{0.05in}}
  \hspace{1in}
  Erin Hartman\thanks{Assistant Professor, Department of Statistics
    and Department of Political Science, University of California, Los Angeles, Los Angeles, CA 90095. Email:
    \href{mailto:ekhartman@ucla.edu}{ekhartman@ucla.edu}, URL: \url{www.erinhartman.com}}
}

\date{This Draft: December 19, 2020}

\maketitle

% }\fi

%   \if1\blind

% \maketitle
% \fi

\spacingset{1.25}
\pdfbookmark[1]{Title Page}{Title Page}

\thispagestyle{empty}
\setcounter{page}{0}
\begin{abstract}
  Generalizing estimates of causal effects from an experiment to a
  target population is of interest to scientists. However, researchers
  are usually constrained by available covariate information. Analysts can often collect much fewer variables from population
  samples than from experimental samples, which has limited applicability of existing approaches that
  assume rich covariate data from both experimental and population
  samples. In this article, we examine how to select covariates
  necessary for generalizing experimental results under such data
  constraints. In our concrete context of a large-scale
  development program in Uganda, although more than 40 pre-treatment covariates are available in the experiment, only 8 of them were also
  measured in a target population. We propose a method to estimate a {\it separating} set -- a set of variables affecting both the sampling mechanism and
  treatment effect heterogeneity -- and show that the population
  average treatment effect (PATE) can be
  identified by adjusting for estimated separating sets. Our
  algorithm only requires a rich set of covariates in the
  experimental data, not in the target population, by incorporating researcher-specific constraints on what variables are measured in the population data.
  Analyzing the development experiment in Uganda, we show that the proposed algorithm can allow for the PATE estimation in situations where
  conventional methods fail due to data requirements.
\end{abstract}

\noindent%
\vspace{0.2in}
\vfill

\newpage
\newcommand\origspace{1.5}
\spacingset{\origspace}
\section{Introduction}
Over the last few decades, social and biomedical scientists have developed and applied an array of statistical tools to make valid
causal inferences \citep{imbens2015causal}. In particular, randomized experiments have become
the mainstay for estimating causal effects. Although many
scholars agree upon the high internal validity of experimental
results, there is a debate about how scientists should infer the impact of policies and interventions
on broader populations \citep{imai2008misunderstandings,
  Angrist:2010jv, Imbens:2010hu, bare2016causal, Deaton:2017kg}.  This issue of
generalizability \citep{Stuart:2011hr} is pervasive in
practice because randomized controlled trials are often conducted on
non-representative samples \citep{shadish2002experiment,
  druckman2011cambridge, Allcott:2015eh,
  Stuart:2015id}.

In this paper, we examine how to generalize the experimental results of the Youth Opportunities Program (YOP) in
Uganda, which aims to help the poor and unemployed become self-employed
artisans and increase incomes. This large scale development program,
involving more than 10,000 individuals was implemented by the government of Uganda and the authors
of \citet{blattman2013generating} from 2008 to 2012. Young adults in
Northern Uganda were invited to form groups and submit grant proposals
for vocational training and to start independent trades. To evaluate
the causal impact of the program, funding was randomly assigned and a host of economic variables (e.g.,
employment and income) were measured. %See Section~\ref{sec:app} for
%the details of the experiment and Section~\ref{sec:app_ana} for our
%empirical analysis.

The question of generalizability is especially important in this application. The aim
of such development programs is elegantly noted in \citet{duflo2005use}, ``the benefits of knowing which
programs work and which do not extend far beyond any program or agency,
and credible impact evaluations are global public goods in the sense
that they can offer reliable guidance to international organizations,
governments, donors, and nongovernmental organizations (NGOs) beyond
national borders.'' Researchers and policy makers are not just concerned to learn about
the very individuals who participated in the trial. The
ultimate goal is to learn whether and how much the program can improve
economic conditions in a larger target population --- about 10
million people in Northern Uganda \citep{uganda2007}.

Despite its importance, estimating population average treatment
effects is not straightforward because we have
to adjust for differences between experimental samples and the target
population. One pervasive question is what covariates should and can we adjust for? Although previous
research shows that adjusting for a set of variables explaining
sampling mechanism or treatment heterogeneity is sufficient for
generalization \citep{Stuart:2011hr, bare2014recover}, researchers are often constrained by available
covariate information in applied settings.

In this paper, we address this problem of covariate selection for
estimating population average treatment effects. In particular, we
develop a data-driven method to estimate {\it a separating set} -- a set of variables affecting
both sampling mechanism and treatment effect heterogeneity. Recent
papers show that the population average treatment effect can be identified by adjusting for
this separating set \citep{Cole:2010bf, Tipton:2013ew, Pearl:2014hb,
  Kern:2016ez}. In Section~\ref{sec:sep}, we extend this result and show that the separating set
relaxes data requirements of conventional methods by generalizing two
widely-used covariate selection approaches: (1) {\it a sampling set} -- a set of variables
explaining how units are sampled into a
given experiment \citep{pressler2013use, Hartman:2015hq, Buchanan:2018dd} and (2)
{\it a heterogeneity set}: a set of variables explaining treatment
effect heterogeneity \citep{Kern:2016ez, Nguyen:2017jw}.

In Section~\ref{sec:dis}, we demonstrate that such separating sets are estimable from the
experimental data and provide a new estimation algorithm based on
Markov random fields. This algorithm only requires that a sampling set be observed
in the experimental sample, not in the target population. We estimate
a separating set as a set that makes a sampling set conditionally independent
of observed outcomes in the experimental data. Therefore,
in contrast to conventional methods, we can exploit all covariates in the experiment to find
necessary separating sets, even when there are few variables measured
in both the experimental and population data.

Importantly, our proposed approach maintains a widely used
assumption that a sampling set is observed in the experimental
data. However, unlike many existing methods, we do not assume that a
sampling set is also observed in the population data. This distinction in data
requirement is subtle and yet practically essential because in many applied contexts, a larger number of
covariates are measured in the experimental data than in the population data. For example, the experimental data of
\citet{blattman2013generating} contains about 40 pre-treatment
covariates, even though only 8 of them are also measured in the target
population. To estimate separating sets, our proposed method
incorporates such user-constraints
on what variables can feasibly be collected in the target
population.  For instance, suppose people selected into the YOP
due to social connections, which were unmeasured in the target
population. Even in this scenario where conventional methods fail,
the proposed method can estimate separating sets accounting for this data constraint, if any exist.

% Our approach has two advantages. First, unlike existing approaches
% requiring rich covariate information in both the experimental sample
% and the non-experimental target population, we only require rich
% covariate data in the experimental sample. This is practically
% important because researchers often have more control over
% what to measure on their experimental subjects, even when it is
% difficult to collect detailed information about their target
% population. % The second related advantage is that our covariate
% % selection algorithm can be implemented even before collecting the
% % target population data. The proposed algorithm can inform
% % which types of variables they should measure in the target
% % population.
% Second, we can incorporate user-constraints
% on what variables can feasibly be collected in the target population.
% If there are characteristics that cannot be measured in the target
% population, the algorithm will identify a separating set subject to
% these constraints. %For instance, suppose people selected into the YOP
% %due to social connections, which were hard to measure in the target
% %population. Even in this scenario, we can estimate separating sets
% %accounting for this data constraint, if any exist.

Our article builds on a growing literature on the population average
treatment effect, which has two general directions. First, many previous studies have focused on articulating identification
assumptions and proposing consistent estimators of the population
average treatment effect \citep[e.g.,][]{Stuart:2011hr,
  Hartman:2015hq, Buchanan:2018dd}. In particular, recent papers
explicitly show that researchers have to jointly consider
treatment effect heterogeneity and the sampling mechanism \citep{Cole:2010bf, Tipton:2013ew, Pearl:2014hb,
  Kern:2016ez}. These existing approaches often assume researchers have access to a large number
of covariates in both the experimental sample and the non-experimental
target population. In contrast, we provide a new data-driven
covariate selection algorithm to find separating sets in
situations where researchers have data constraints in the target
population. Our focus on covariate selection is similar to recent
influential work on causal directed acyclic graphs (causal DAGs)
\citep{bare2014recover, bare2016causal}. We differ from the DAG-based
approaches in that we empirically estimate separating sets under assumptions
about sampling and heterogeneity sets rather than analytically selecting separating
sets from fully specified causal DAGs. Although assumptions about
the entire causal DAGs are sufficient for covariate
selection, the proposed algorithm can estimate separating sets under
weaker assumptions about sampling and heterogeneity sets at the expense of statistical uncertainty.

Research in the second direction argues that the necessary assumptions
for existing methods are often too strong in practice. Recent papers
have explored methods for sensitivity analyses \citep[e.g.,][]{andrews2019} and bounds \citep[e.g.,][]{Chan:2017de} to achieve partial
identification under weaker assumptions. In particular, \citet{Nguyen:2017jw,
  nguyen2018} also consider data scenarios where the experimental data
has a richer set of covariates than the target population data, and propose
a sensitivity analysis by specifying sensitivity parameters that
captures the distribution of covariates unmeasured in the target
population data. Our paper is complementary to these approaches. We instead focus on the point identification of the
population average treatment effect and alleviate strong assumptions
about data requirements by adding an additional step of estimating a
separating set. Our approach can also be used in conjunction with
sensitivity analysis; the proposed method can estimate a smaller
separating set, thereby reducing the number of unobserved covariates that sensitivity analyses have to consider. We
provide further discussions about relationships between our
proposed approach and sensitivity analysis in Section~\ref{subsec:sa}.

\section{Youth Opportunities Program in Uganda}
\label{sec:app}
As well documented by the World Bank, a large number of young adults in
developing countries are unemployed or underemployed \citep{worldbank2012}. In addition to its direct implication to poverty,
concerns for policy makers are that such large young and
unemployed populations can increase risk of crime and social unrest
\citep{blattman2013generating}. Uganda,
especially conflict-affected Northern Uganda, is not an exception. According to
estimates from the government, two-thirds of northern
Ugandans could not meet basic needs, about 50\% were illiterate, and most
were underemployed in subsistence agriculture in 2006 \citep{uganda2007}.

In this paper, we study the Youth Opportunities Program (YOP) in
Uganda, designed to help the poor and unemployed become self-employed
artisans and increase incomes. This intervention is one example of widely used cash
transfer programs in which participants are offered a certain amount of
cash in the hope that they invest in training and start new,
profitable enterprises. In 2008, the government invited young adults in
Northern Uganda to form groups and submit grant proposals for how they
would use a grant for vocational training and business start-up. Then,
funding was randomly assigned among 535 screened, eligible applicant
groups --- 265 and 270 groups to treatment and control, respectively. Treatment groups received a
one-time unsupervised grant worth \$7,500 on average --- about \$382
per group member, roughly their average annual income. Following the
original analysis, we focus on a binary treatment, whether they receive any grants or not through the YOP.

To evaluate the impact of this intervention,
\citet{blattman2013generating} surveyed 5 people per  group three times
over four years, resulting in a panel of 2,598 individuals after removing 79 observations due to missing data. They measured 17
 outcome variables across five dimensions ---
employment (7), income (2), investments (3),  business formality (3),
and urbanization (2). They find that the effects of the YOP are large across
all dimensions. Notably, after two years, the treatment groups were 4.5 times more likely to
have vocational training, 2.6 times more likely to engage with a skilled
trade and had 16\% more hours of employment and 42\% higher earnings.

\newcommand{\PreserveBackslash}[1]{\let\temp=\\#1\let\\=\temp}
\newcolumntype{C}[1]{>{\PreserveBackslash\centering}p{#1}}
\newcolumntype{R}[1]{>{\PreserveBackslash\raggedleft}p{#1}}
\newcolumntype{L}[1]{>{\PreserveBackslash\raggedright}p{#1}}

Although it is unambiguous that the YOP had large, persistent positive
effects on experimental subjects, it is of great policy interest to
empirically investigate how much these experimental estimates are
generalizable to a larger population. Estimating population average
treatment effects (PATE) can inform which specific development policies governments
should scale up. While the focus of the program was on Northern Uganda as
a whole, participants of the YOP were inevitably not representative, as in many other development
programs. To take into account differences between experimental
samples and Northern Uganda's population, \citet{blattman2013generating} merged their
experimental samples with a 2008 population-based
household survey, the Northern Uganda Survey (NUS). They adjusted for eight variables shared by
experimental and population data; gender, age, urban status, marital
status, school attainment, household size, durable assets, and district
indicators.

In Table~\ref{tab:app}, we report estimates based on an inverse probability
weighting (IPW) estimator \citep{Stuart:2011hr} that adjusts for the
original eight variables.\footnote{\spacingset{1}{\footnotesize Although the original authors rely
  on weighted linear regression models in their paper, we focus on the
  IPW estimator widely studied in the literature of generalization
  \citep[e.g.,][]{Buchanan:2018dd}.}} As a reference, we also report estimates of
the average treatment effect within the experimental sample, called
the sample average treatment effect (SATE). Estimates of the
SATEs and PATEs have roughly the same sign, suggesting that the
program will have a positive impact on a variety of outcomes even in
the target population. Importantly, this finding, however, rests on an assumption
that the original eight variables adjust for all relevant differences
between the experimental sample and the target population. Given
that the magnitude of the PATE estimates has strong implications for
a cost-benefit analysis of these large-scale expensive interventions,
it is critical to examine this common methodological challenge of
covariate selection for generalizing experimental results.

% their exact magnitude
% often differs, which is critical for a cost-benefit analysis of these
% types of large-scale expensive interventions. For example, compared to the SATE, the PATE is 46\%
% larger on durable assets, 58\% larger on records maintenance, and yet
% 28\% smaller on business assets, although these differences are not
% statistically significant due to larger standard errors of the PATE estimate. Importantly, the statistically insignificant difference
% between the SATE and the original PATE estimate does not imply that the intervention will perform as well as in the target
% population. This is because it is difficult to know whether the
% original eight control variables adjust for all relevant differences
% between the experimental samples and the target population of interest.

\begin{table}[!t] \centering
  \small
  \resizebox{\textwidth}{!}{
    \begin{tabular}{@{\extracolsep{5pt}} L{2.1in}C{0.5in}C{0.55in}|L{2in}C{0.5in}C{0.55in}}
      \hline  \hline
      & & Original &  & & Original  \\[-5pt]
      & SATE & PATE &  & SATE & PATE  \\[-8pt]
      & estimate & estimate &  & estimate & estimate \\ \hline
      \normalsize \bf \underline {Employment} &  &  &  \normalsize \bf \underline {Investments} &  &  \\
      Average employment hours & 5.13 & 5.87 & Vocational training & 0.55 & 0.51 \\[-3pt]
      & (1.22) & (3.25) &  & (0.03) & (0.07) \\
      Agricultural & -0.01 & 0.75 & Hours of vocational training & 349.65 & 277.18 \\[-3pt]
      & (1.02) & (2.01) &  & (23.61) & (51.21) \\
      Nonagricultural & 5.14 & 5.12 & Business assets & 426.17 & 400.31 \\[-3pt]
      & (0.92) & (2.62) &  & (82.79) & (125.19) \\
      Skilled trades only & 4.75 & 2.73 &  &  &  \\[-3pt]
      & (0.64) & (1.56) &  \normalsize \bf \underline {Business Formality} &  &  \\
      No employment hours & -0.02 & 0.00 & Maintain records & 0.12 & 0.19 \\[-3pt]
      & (0.01) & (0.02) &  & (0.03) & (0.07) \\
      Any skilled trade & 0.27 & 0.24 & Registered & 0.06 & 0.07 \\[-3pt]
      & (0.03) & (0.07) &  & (0.02) & (0.04) \\
      Works mostly in a skilled trade & 0.06 & -0.01 & Pays taxes & 0.08 & -0.01 \\[-3pt]
      & (0.01) & (0.03) &  & (0.02) & (0.06) \\
       \normalsize \bf \underline {Income} &  &  &  \normalsize \bf \underline {Urbanization} &  &  \\
      Cash earnings & 13.41 & 12.15 & Changed parish & 0.01 & -0.03 \\[-3pt]
      & (3.87) & (5.3) &  & (0.02) & (0.07) \\
      Durable assets & 0.11 & 0.09 & Lives in Urban area & -0.01 & -0.06 \\[-3pt]
      & (0.05) & (0.16) &  & (0.03) & (0.07) \\
      \hline
    \end{tabular}}
  \spacingset{1.2}{
    \caption{Estimates of Sample Average Treatment Effects and
      Population Average Treatment Effects based on
      the Original Eight Variables. {\it Note:} We estimated
      population average treatment effects (PATE) of the above 17 outcomes
      using an inverse probability weighting
      estimator with standard errors clustered by group. Weights are estimated by a logistic
      regression including the eight variables additively. See details
      of the estimation in Section~\ref{sec:app_ana}. As a reference, we
      also report estimates of the sample average treatment effect (SATE).}\label{tab:app}}
\end{table}

In practice, there are several pervasive concerns about covariate selection. First, although it is common to adjust
for all observed covariates shared by experimental and population
data, it is unclear whether such sets of covariates include all
necessary covariates for generalization. In fact, the authors
carefully pay attention to this point in the original paper; ``young adults are
selected into our sample because of unobserved initiative, connections
or affinity for entrepreneurship'' \citep{blattman2013generating}. If
there are unobserved differences between the experimental and
population samples, the original PATE estimate would be
biased. Second, it is also possible that the original analysis adjusted for unnecessary
variables, resulting in inefficient estimators of the PATE.
\citet{miratrix_sekhon_theodoridis_campos_2018} show that weighting on
many variables, particularly those not highly correlated to treatment effect
heterogeneity, can lead to inefficient estimation of the PATE. In this paper, we investigate necessary and sufficient sets of
covariates for generalizing experimental estimates, called separating sets, and then
provide a new algorithm to empirically estimate such sets. We select
the separating sets under several different assumptions and assess how
estimates of the PATE vary. Our reanalysis of this experiment appears in Section~\ref{sec:app_ana}.

\section{Separating Sets For Generalization}
\label{sec:sep}
This section sets up the potential outcomes framework
\citep{neyman1923, rubin1974causal} for
studying population average treatment effects. We review a definition
of {\it a separating set} --- a set of variables affecting both the sampling
mechanism and treatment effect heterogeneity, and then show that a sampling set and a heterogeneity set, the
main focus of existing approaches, are special cases of the separating
sets. % Finally, we demonstrate how we can identify separating sets just
% from experimental data.  Estimation of separating sets and population
% average treatment effects is considered in the next section.

\subsection{The Setup}
\label{subsec:setup}
We consider a scenario in which we have two data sets. Following \citet{Buchanan:2018dd}, we define the
first sample of $n$ individuals to be participants in a randomized experiment
(``Experimental Data'') and the second data set to be a random sample of $m$
individuals from the target population (``Population Data''). In our
application, the experimental data has 2,598 individuals and the population
data contains 21,348 individuals. We define a sampling indicator $S_i$ taking
$1$ if unit $i$ is in the experiment and $0$ if unit $i$ is in the target population. We assume that every unit has non-zero
probability of being in the experiment.\footnote{This assumption of
  non-zero sampling probability is known as the ``Positivity of trial
  participation'' \citep{colnet2020causal}, and
  is commonly made in the generalization literature. This assumption is untestable and
  researchers have to evaluate it with domain knowledge
  \citep{dahabreh2020extending}. When this assumption is violated,
  researchers have to rely on modeling assumptions and model-based
  extrapolation \citep{mealli2019overlap}. Alternatively, researchers
  can restrict their attention to a subset of the target population
  that has non-zero sampling probability \citep[e.g.,][]{tipton2016site}, which has been the most
  common approach in the causal inference literature to deal with
  non-overlap between the treatment and control groups.} Although experimental
units can be randomly sampled from the target population in ideal
settings, units often non-randomly select into the experiment, as in the YOP, making the
experimental sample non-representative. Note that we consider cases in which
units are either in the experimental data or in the target population data,
but similar results hold for cases in which the experimental sample is
a subset of the target population.

Let $T_i$ be a binary treatment assignment variable for unit $i$
with $T_i= 1$ for treatment and $0$ for control. We define $Y_i(t)$ to be the potential outcome variable of unit $i$ if the unit
were to receive the treatment $t$ for $t \in \{0, 1\}$. In this paper,
we make a stability assumption, which states that there is neither
interference between units nor different versions of the treatment, either across units or settings
\citep{Rubin:1990tm, Tipton:2013ew, Hartman:2015hq}. We define pre-treatment
covariates $\bX_i$ to be any variables not affected by the treatment
variable.
% Researchers are rarely able to conduct their experiment on the target
% population itself, and therefore often conduct an randomized
% experiment with a different population.

We are interested in estimating the average treatment effect in the
target population. We call this causal estimand the population average
treatment effect (PATE).
\vspace{0.1in}
\begin{definition}[Population Average Treatment Effect]
    \ \vspace{-0.2in}
    \spacingset{1.2}{
     $$ \tau \equiv \E[Y_i(1) - Y_i(0) \mid S_i = 0], $$
     where $S_i = 0$ represents the target population data.
}
\end{definition}\vspace{-0.1in}
The treatment assignment mechanism is controlled by researchers within
the experiment ($S_i=1$), but it is unknown for units in the target
population ($S_i=0$; observational data). Formally, we assume that the treatment assignment is randomized within the experiment.
\vspace{0.1in}
\begin{assumption}[Randomization in Experiment]
  \label{random}
  \ \vspace{-0.2in}
  \spacingset{1}{
  $$  \{Y_i(1), Y_i(0), \bX_i\} \ \indep \ T_i \mid S_i=1$$}
\end{assumption}\vspace{-0.1in}
This assumption holds by design in randomized experiments. Here, we
consider unconditional randomization, but results in the paper can be
naturally extended to settings with randomization conditional on some
pre-treatment covariates. Finally, for each unit in the experimental
condition, only one of the potential outcome variables can be
observed, and the realized outcome variable for unit $i$ is denoted by
$Y_i = T_i Y_i(1) + (1-T_i) Y_i(0)$ \citep{rubin1974causal}.

%While both outcomes and treatments need to be measured in the experimental data, we do not require information about treatments or outcomes for the non-experimental target population.

\subsection{Definition of Separating Sets and Identification}
\label{subsec:sep_set_def}
Recent papers show that the PATE can be identified by a set of
variables affecting both treatment effect heterogeneity and the
sampling mechanism \citep{Cole:2010bf, Tipton:2013ew, Pearl:2014hb,
  Kern:2016ez}. In this paper, we refer to this set as a {\it separating set} and
investigate its statistical properties. Formally, a separating set is
any set that makes the sampling indicator and treatment effect
heterogeneity conditionally independent.
\vspace{0.1in}
\begin{definition}[Separating Set]
  \label{sep}
    A separating set is a set $\bW$ that makes the sampling indicator and treatment effect
    heterogeneity conditionally independent.
  \begin{equation}
  Y_i(1) - Y_i(0) \ \indep \ S_i \mid \bW_i.\label{eq:sepeq}
  \end{equation}
\end{definition}\vspace{-0.1in}
This definition of a separating set contains two simple cases: (1)
when no treatment effect heterogeneity exists and (2) when the
experimental sample is randomly drawn from the target population. In
both of these cases, $\bW_i = \{\varnothing\}$. This separating set
also encompasses two common approaches in the literature as special
cases. First, researchers often employ statistical methods based on a
\emph{sampling set} -- a set of all variables affecting the sampling
mechanism \citep[e.g.,][]{Stuart:2011hr}. Second, researchers might adjust for a \emph{heterogeneity
  set} -- a set of all variables governing treatment effect
heterogeneity \citep[e.g.,][]{Kern:2016ez}. Below, we formalize these sets based on the potential
outcomes framework.

We define a {\it sampling set} as a set of variables that determines
the sampling mechanism by which individuals come to be in the
experimental sample. For example, when a researcher implements
stratified sampling based on gender and age, the sampling set consists of
those two variables. When researchers control the sampling mechanism,
a sampling set is known by design. However, when samples are selected
without such an explicit sampling design, a sampling set is unknown
and in practice, researchers must posit a sampling mechanism.  % Often, researchers will adjust for all variables measured in both the experimental data and target population, a set which may or may not meet the conditional independence statement in Equation~\eqref{eq:sepeq}.
For example, \citet{blattman2013generating} assume that a sampling set
consists of eight variables: gender, age, urban status, marital
status, school attainment, household size, durable assets, and district
indicators.

Formally, we can define a sampling set $\bX^S$ as follows.
\vspace{0.1in}
\begin{definition}[Sampling Set]
  \label{sample}
  \spacingset{1.2}{ $ $ \vspace{-0.35in}\\
    \begin{equation}
      \{Y_i(1), Y_i(0), \bX_i^{-S}\} \ \indep \ S_i \mid \bX_i^S\label{eq:par}
    \end{equation}}
  where $\bX^{-S}$ is a set of pre-treatment variables that are not in
  $\bX^S$.
\end{definition}
This conditional independence means that the sampling set
is a set that sufficiently explains the sampling mechanism. Given the
sampling set, the sampling indicator is independent of the joint
distribution of potential outcomes and all other pre-treatment
covariates. We refer to variables in the sampling set as sampling
variables.
% \footnote{Note that one advantage here is that the sampling
  % set yields conditional independence of the joint distribution of
  % potential outcomes, rather than the difference, which allows for
  % generalization of a wider range of estimands, not just the
  % population average treatment effect defined as the difference in the
  % potential outcomes.}

The other popular approach is to adjust for a set of all variables
explaining treatment effect heterogeneity, which we call a {\it
  heterogeneity set}. Formally, we can define a heterogeneity set
$\bX^H$ as follows.
\vspace{0.1in}
\begin{definition}[Heterogeneity Set]
  \label{hetero}
  \spacingset{1.2}{ $ $ \vspace{-0.35in}\\
    \begin{equation}
  Y_i(1) - Y_i(0) \ \indep \  \{S_i, \bX_i^{-H}\} \mid \bX_i^H,
    \end{equation}}
where $\bX^{-H}$ is a set of pre-treatment variables that are not in
$\bX^H$.
\end{definition}
In this case, because a heterogeneity set fully accounts
for treatment heterogeneity, $Y_i(1) - Y_i(0)$ is independent of all
other variables. We refer to variables in the heterogeneity set as
heterogeneity variables. In our application,
\citet{blattman2013generating} discuss at least two heterogeneity
variables, gender and initial credit constraints.

We want to emphasize that a sampling set and a heterogeneity set are special cases of a
separating set in the sense that both sets satisfy
equation~\eqref{eq:sepeq}. Yet, there may exist many other separating
sets, which we explore in Section~\ref{sec:dis}.

Finally, the PATE is nonparametrically identified by adjusting for a separating set \citep{Cole:2010bf, Tipton:2013ew, Pearl:2014hb,
  Kern:2016ez}.
\begin{result}[Identification of the PATE]
\label{pate} The PATE is identified with separating set $\bW_i$ under Assumption~\ref{random}.
  \begin{eqnarray*}
    \tau  & = & \int \biggl\{\E[Y_i \mid T_i =1, S_i =1, \bW_i=\bw] - \E[Y_i \mid T_i =0, S_i =1, \bW_i=\bw] \biggr\} d F_{\bW_i \mid S_i=0}(\bw),
  \end{eqnarray*}
  where $F_{\bW_i \mid S_i = 0}(\bw)$ is the cumulative distribution function of $\bW$ conditional on $S_i=0$.
\end{result}
As sampling and heterogeneity sets are special cases of a separating
set, the PATE is identified with the same formula.

\subsubsection{Illustration with A Causal DAG.}
\begin{figure}[!t]
  \begin{center}
    \includegraphics[width = 0.5\textwidth]{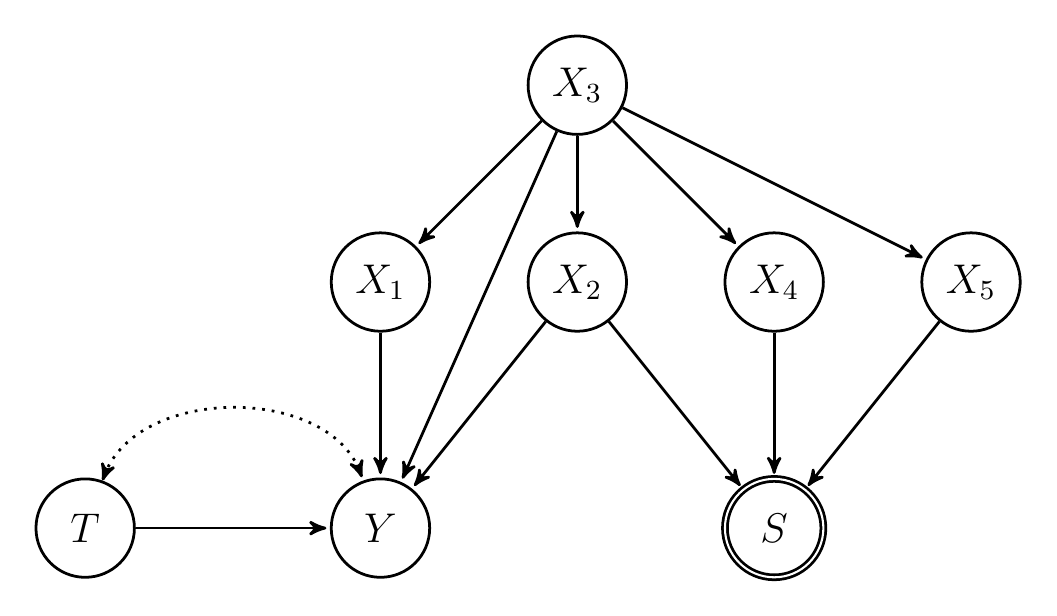}
    \spacingset{1.2}{\caption{Example of a Causal DAG based on a
        selection diagram approach
        \citep{bare2016causal}. \textit{Note:} Three variables
        $\{X_2, X_4, X_5\}$ serve as a sampling set, three variables
        $\{X_1, X_2, X_3\}$ as a heterogeneity set, and two variables
        as the smallest separating set $\{X_2, X_3\}$. }\label{fig:dag}}
  \end{center}
\end{figure}

We consider a causal DAG in Figure~\ref{fig:dag} as a concrete
illustration based on a selection diagram approach
\citep{bare2016causal}. In this causal DAG, three variables
$\{X_2, X_4, X_5\}$ serve as a sampling set, which has direct arrows
to sampling variable $S$. Three variables $\{X_1, X_2, X_3\}$ serve as
a heterogeneity set, which has direct arrows to outcome variable $Y$
and can moderate the causal effect of $T$ on $Y$. Finally, from
Definition~\ref{sep}, there are many valid separating sets, including
the sampling and heterogeneity sets, but the smallest separating set
is $\{X_2, X_3\}$, which makes the sampling indicator $S$ and the
outcome variable $Y$ conditionally independent. The key is that there
are potentially many valid separating sets, and researchers can use
any of them for identification given their data constraint.

\begin{remark}
  We only use the causal DAG in Figure~\ref{fig:dag} as an
  illustrative example, and our
  proposed approach relies only on assumptions we clarify in
  Section~\ref{sec:dis} using the potential outcomes. We do
  not use any particular causal DAG structure in
  Figure~\ref{fig:dag}, and we do not assume knowledge of the underlying causal DAG structure. When
  knowledge of the underlying causal DAG is available, results in the
  causal diagram literature are of great importance, and we refer
  readers to \citet{bare2016causal}.
\end{remark}

\section{Identification and Estimation of Separating Sets}
 \label{sec:dis}
% {\bf \color{red} NE: This Motivation part should be better.}
% A separating set is useful not only as a conceptual generalization of
% the two well-known existing approaches --- sampling and
% heterogeneity sets --- but also as a way to incorporate data
% constraints. An under-explored advantage of
% separating sets over sampling and heterogeneity
% sets is that there may exist, potentially many, other separating
% sets and hence, researchers can choose a set subject to their data
% constraints. For example, researchers might be able to measure some sampling variables, e.g.,  social connections and
% initial motivation for entrepreneurship, only in the experimental data
% and not in the target population \citep{blattman2013generating}.

 In this section, we first show that a variant of {\it separating} sets,
 which is sufficient for the identification of the PATE, is estimable
 even when a {\it sampling} set is unobserved in the population data
 as far as it is observed in the experimental data (Section~\ref{subsec:iden}). This is in contrast to existing methodologies
 which assume that a sampling set is observed in {\it both} of the
 experimental and population data. This distinction is subtle and yet
 practically important because in many
 applied contexts, including the YOP \citep{blattman2013generating}, a larger number of covariates are
 measured in the experimental data than in the population data. This
 is because, while analysts of experiments can often control what variables should be measured within the experiment, population data
 is usually more expensive; collected by other organizations, such as a national-level survey (the NUS in
 \citet{blattman2013generating}); or otherwise impractical to
 collect. Thus, our focus is on this type of common research setting where
 analysts are able to measure more covariates in the experimental data
 than in the population data.

After demonstrating the identification of separating sets, we propose an algorithm to estimate
separating sets using Markov random fields (Section~\ref{subsec:est}).

\subsection{Identification of Separating Sets}
\label{subsec:iden}
We begin with the identification of an exact separating set and then
turn to the identification of a modified separating set under a weaker assumption.

First, we can estimate an exact separating set in settings where both a sampling set and a
heterogeneity set are observed in the experimental data. A key feature
of this result is that we only require rich covariate information about the experimental units, not the target population
units, to discover separating sets, should they exist.

In many applied research contexts, however, the heterogeneity set is
not readily available even in the experimental data. The fundamental
problem of causal inference states that only one of two potential outcomes
are observable, which implies that the causal effect is unobserved at
the unit level, and thus so is the heterogeneity set. For example, in our
application, although \citet{blattman2013generating} discuss two
specific heterogeneity variables (gender and initial credit
constraints), it might be unreasonable to assume away the existence of
other potential heterogeneity variables.

We therefore develop an additional method to find a variant of a separating set, which we call
a {\it marginal} separating set, using only knowledge of a sampling set, a commonly employed assumption in the extant literature.
We show that a marginal separating set can be discovered when
a sampling set is measured in the experimental data, but not in the target
population. Although this data requirement
might still be stringent in some contexts, it is much weaker than the
one necessary for widely-used existing approaches based on sampling
sets, which require that sampling set be measured in the population
data as well as in the experimental data.
%on a sampling set or a heterogeneity set.

\subsubsection{Identification of Exact Separating Sets}\label{subsubsec:disc_sep_set}
We begin with settings in which a sampling set and a heterogeneity set
are observed in the experimental sample. In this setting, we can use
the experimental data to identify exact separating sets. Although this
data requirement is still restrictive, we emphasize that it does not
require rich data on the target population.
\vspace{0.15in}
\begin{setting}[Sampling and Heterogeneity Sets are Observed in Experiment]
  \label{obsSamHet} \spacingset{1}{
    Sampling set $\bX^S$ and heterogeneity set $\bX^H$ are observed in the experiment ($S_i=1$).}
\end{setting}
\vspace{-0.05in}
In this setting, a separating set is estimable as a set that makes the
sampling set and the heterogeneity set conditionally independent
within the experimental data.
\vspace{0.05in}
\begin{theorem}[Identification of Separating Sets in Experiment]
\label{thm_id_Exsep}
  \vspace{0.1in} \spacingset{1}{
    In Setting~\ref{obsSamHet}, consider a set of covariates $\bW$, which is a subset of
    pre-treatment variables. Under Assumption~\ref{random},
    \begin{eqnarray}
      \widetilde{\bX_i}^H \ \indep \ \widetilde{\bX_i}^S \mid \bW_i, T_i,
      S_i=1 & \Longrightarrow &  Y_i (1) - Y_i(0) \ \indep \  S_i \mid
                                \bW_i, \label{eq:Exsep}
    \end{eqnarray}
    \vspace{-0.1in}\\
    where $ \widetilde{\bX}^H$ and $ \widetilde{\bX}^S$ are the set difference $\bX^H \setminus \bW$ and $\bX^S \setminus \bW$, respectively.}
\end{theorem}
We provide the proof in the supplementary material. Theorem~\ref{thm_id_Exsep} states that as long as we
can find a set that satisfies the testable conditional independence on
the left hand side, the discovered set is guaranteed to be a
separating set. That is, we can identify an exact separating set from
the experimental data alone. Note that when $\bX^H$ and $\bX^S$
share some variables, those variables should always be in
$\bW$. Using the selected separating set, researchers can identify the
PATE based on Result~\ref{pate}.

Intuitively, this theorem can be explained through two conceptual steps. First, because a
heterogeneity set $\bX^H$ fully explains treatment effect
heterogeneity $Y(1) - Y(0)$, the sampling indicator $S$ and $Y(1) - Y(0)$ are conditionally dependent only when
$S$ and $\bX^H$ are conditionally dependent. Second, because
a sampling set $\bX^S$ fully explains the sampling indicator $S$,
$S$ and $\bX^H$ are conditionally dependent only when $\bX^S$
and $\bX^H$ are conditionally dependent. Taken together, $S$ and
$Y(1)- Y(0)$ are conditionally dependent only when $\bX^S$ and $\bX^H$
are conditionally dependent.

\paragraph{Example.}
Here, we illustrate the general result from Theorem~\ref{thm_id_Exsep} using
the causal DAG in Figure~\ref{fig:dag} as a concrete example. Suppose we are
interested in a separating set $\bW = \{X_2, X_3\}$, which has the
smallest size. Given that heterogeneity set $\bX^H = \{X_1, X_2,
X_3\}$ and $\bX^S = \{X_2, X_4, X_5\}$, we have $\widetilde{\bX}^H =
X_1$ and $\widetilde{\bX}^S = \{X_4, X_5\}.$ Then, as
equation~\eqref{eq:Exsep} suggests, we have $X_1 \ \indep \ \{X_4, X_5\}
\mid X_2, X_3, T, S = 1$ in the causal DAG in Figure~\ref{fig:dag}. More generally, Theorem~\ref{thm_id_Exsep}
guarantees that any set that makes the sampling set and the
heterogeneity set conditionally independent within the experimental
data is a separating set under Assumption~\ref{random}.

\subsubsection{Identification of Marginal Separating Sets}
\label{subsubsec:marginal_sep_sets}
While Theorem~\ref{thm_id_Exsep} allows us to discover separating sets
using the experimental data, a key challenge would be to measure both
a sampling set and a heterogeneity set in the experimental data.  In
particular, it is often difficult to measure the heterogeneity set in
practice. We show that a
modified version of a separating set -- a {\it marginal} separating
set -- is estimable from the experimental data under a weaker assumption. We define a marginal separating set as follows.
\begin{definition}[Marginal Separating Set]
  \label{marginal_sep}
  \vspace{0.1in} \spacingset{1}{
    A marginal separating set is a set $\bW$ that makes the sampling
    indicator and the marginal distributions of potential outcomes conditionally independent.
  \begin{equation}
      Y_i(t) \ \indep \ S_i \mid \bW_i \hspace{0.2in} \mbox{for  } t = \{0, 1\}.\label{eq:marginal_sep}
  \end{equation}}
\end{definition}\vspace{-0.1in}
We refer to this as a {\it marginal} separating set since it renders the
marginal, not the joint, distribution of potential outcomes
conditionally independent of the sampling process.

Now we turn to our final setting researchers may find themselves in -- that the
sampling set is observed only in the experimental data.  Previous work
using the sampling set assumes it is measured in both the experimental
sample and the target population \citep[e.g.,][]{Cole:2010bf,
  Tipton:2013ew, Hartman:2015hq, Buchanan:2018dd}.  Since researchers often have much
more control over what data is collected in the experiment, this final
setting greatly relaxes the data requirements of the previous
literature. %When we measure the sampling set in the experimental data, we can identify a marginal separating set as a set that makes the sampling set and the observed outcomes conditionally independent within the experimental data.

\begin{setting}[Sampling Set is Observed in Experiment]  \label{obsSamEx}
  \label{obsSam}
\vspace{0.1in} \spacingset{1}{
  Sampling set $\bX^S$ is observed in the experimental data ($S_i=1$).}
\end{setting}

\begin{theorem}[Identification of Marginal Separating Sets in Experiment]
\label{thm_id_sep}
  \vspace{0.1in} \spacingset{1}{
  In Setting~\ref{obsSamEx}, consider a set of covariates $\bW$, which is a
  subset of pre-treatment variables. Under Assumption~\ref{random},
\begin{eqnarray}
 Y_i \ \indep \ \bX_i^S \mid \bW_i, T_i, S_i=1 & \Longrightarrow &  Y_i(t)\ \indep \  S_i \mid \bW_i.
\end{eqnarray}}
\end{theorem}
We provide the proof in the supplementary material. Theorem~\ref{thm_id_sep} states that as long as we
can find a set that makes the observed outcome $Y$ conditionally
independent of the sampling set within the experimental data, the
discovered set is guaranteed to be a marginal separating set. With a large
enough sample size, we can find a marginal separating set from
the experimental data alone. Intuition behind this theorem is similar
to the one used for Theorem~\ref{thm_id_Exsep}. Because the sampling set
$\bX^S$ fully explains the sampling indicator $S$, if the sampling
indicator $S$ and the potential outcome $Y(t)$ are conditionally dependent, the
sampling set $\bX^S$ and the observed outcome $Y$ are also
conditionally dependent.  The marginal separating set may be larger
than an exact separating set, as it may include covariates
that explain the marginal potential outcomes but not treatment effect
heterogeneity. Once we have discovered a marginal separating set using
the experimental data, we can identify the PATE with this discovered set.

\begin{result}[Identification of the PATE with Marginal Separating Sets]
\label{thm_id_Marginal}
  \vspace{0.1in} \spacingset{1}{ $ $\\
  When a marginal separating set $\bW$ is observed both in the
  experimental sample and the target population, the PATE is identified with the marginal separating set $\bW$ under Assumption~\ref{random}.
  \begin{eqnarray*}
    \tau  & = & \int \biggl\{\E[Y_i \mid T_i =1, S_i =1, \bW_i=\bw] - \E[Y_i \mid T_i =0, S_i =1, \bW_i=\bw] \biggr\} d F_{\bW_i \mid S_i=0} (\bw).
  \end{eqnarray*}}
\end{result}
We omit the proof because it is straightforward from the one of Result~\ref{pate}.

\spacingset{1.1}{
\begin{table}[t]
    \label{tab:settings}
    \centering
    \renewcommand{\arraystretch}{1.25}
    \resizebox{\textwidth}{!}{%
    \begin{tabular}{|l|c : c|}
    \hline
      \multirow{2}{*}{\ \ \textbf{Set to Adjust For}} & \multicolumn{2}{c|}{\textbf{Data Requirements}} \\
         & Experiment & Target Population   \\
    \hline \hline
      \  \ Sampling set & Sampling set & Sampling set \\
      \hline
      \ \ Heterogeneity set & Heterogeneity set & Heterogeneity set \\
        \hline
      \begin{tabular}{l}
        Estimated separating set \\
        (Theorem~\ref{thm_id_Exsep} under Setting~\ref{obsSamHet})
      \end{tabular} &  $\left\{\vbox to 12pt{}
                    \begin{tabular}{c}
                            Sampling Set \\
                            Heterogeneity Set
                        \end{tabular} \right.$ & User Specified
                                                 Constraints \\  \hline
      \begin{tabular}{l}
        Estimated marginal separating set\\
        (Theorem~\ref{thm_id_sep} under Setting~\ref{obsSam})
      \end{tabular} &  Sampling set & User Specified Constraints \\
    \hline
    \end{tabular}}
    \caption{Identifying the PATE under
      different data requirements. {\it Note:} Many
      previous approaches assume that a sampling set or a
      heterogeneity set is measured in both the experimental sample
      and the target population (the first two rows). Our proposed approaches relax data requirements for the
      target population by introducing an additional step of estimating separating sets.}
    \label{tab:comparing_assumptions}
\end{table}}
\spacingset{\origspace}

Finally, in Table~\ref{tab:comparing_assumptions}, we compare existing
methods with two proposed approaches. The first two rows show two
common existing approaches based on sampling and heterogeneity sets,
respectively. Although the identification of the PATE in those settings is
straightforward, it requires rich covariate
information from the target population data as well as from the
experimental sample. Our approach relaxes data requirements for the
target population by introducing an additional step of estimating separating sets. In
Setting~\ref{obsSamHet} where we observe both a sampling set and a heterogeneity set
in the experimental sample, we can identify exact separating
sets from the experimental data alone (Theorem~\ref{thm_id_Exsep}). Setting~\ref{obsSam} only requires
observing a sampling set in the experimental sample and we can
identify marginal separating sets (Theorem~\ref{thm_id_sep}). In the
next subsection, we introduce an algorithm that can estimate
separating sets subject to user specified data constraints in the
target population.

\subsection{Estimation of Separating Sets}
\label{subsec:est}
Here, we propose an estimation algorithm to find a marginal separating
set. As shown in Theorem~\ref{thm_id_sep}, the goal is to find a set
that makes a sampling set and observed outcomes conditionally
independent within the experimental data. We show how to apply Markov
random fields (MRFs) to encode conditional independence relationships among observed
covariates and then select a separating set. A similar algorithm can
be used for finding an exact separating set.

Our estimation algorithm consists of four simple steps. We provide a
brief summary here and then describe each step in order. Step 1:
specify all variables in sampling set $\bX^S$ based on domain
knowledge, some of which might not be measured in the population
data. Step 2: using the experimental data alone, estimate a Markov random field over an outcome,
a treatment, the sampling set and observed pre-treatment covariates. Step
3: enumerate all simple paths\footnote{A simple path is a path in a
  Markov graph that does not have repeating nodes.} from $Y$ to
$\bX^S$ in the estimated Markov graph. Step 4: find sets that block
all the simple paths from $Y$ to $\bX^S$ in the estimated Markov graph.

\paragraph{Estimating Markov Random Fields.}
Theorem~\ref{thm_id_sep} implies that we can find a marginal separating
set by estimating a set of variables $\bW$ that satisfies the
conditional independence, $Y_i \ \indep \ \bX_i^S \mid \bW_i, T_i, S=1.$
To estimate this set, we employ a Markov random field (MRF). MRFs are
statistical models that encode the conditional independence structure
over random variables via graph separation rules. For example, suppose
there are three random variables $A, B$ and $C$. Then, $A \indep B
\mid C$ if there is no path connecting $A$ and $B$ when node $C$ is
removed from the graph (i.e., node $C$ {\it separates} nodes $A$ and
$B$), so-called the global Markov property
\citep{lauritzen1996graphical}. We review basic properties of the MRF
in the supplementary material (\ref{sec:mrf-si}), and we refer readers to \citet{lauritzen1996graphical} for
its comprehensive discussion. Using the general theory of MRFs, the
estimation of a separating set can be recast as the problem of finding
a set of covariates separating outcome variable $Y$ and a sampling set
$\bX^S$ in an estimated Markov graph. Therefore, we can find a
separating set that satisfies the desired conditional independence as far as we
can estimate the MRF over $\{Y, T, \bX^S, \bX_0\}$ within the
experimental data where we define $\bX_0$ to be all pre-treatment
variables measured both in the experimental and population data. We
define $\bZ \equiv \{\bX^S, \bX_0\}$ to be pre-treatment covariates
from which we select a separating set.

\begin{remark}
Note that MRFs are used here to estimate
conditional independence relationships of observed covariates as an intermediate step of
estimating separating sets. Importantly, they are not used to estimate the
underlying causal directed acyclic graphs (causal DAGs). As emphasized
earlier in the paper, while we used a causal diagram (Figure~\ref{fig:dag}) as an illustration, we only rely on domain
knowledge about sampling sets, and we are not taking the causal
graphical approach. Such an approach is powerful when knowledge of the
full structure of the underlying causal DAG is available, which we do not assume in this paper.
\end{remark}

%  Failure to estimate a
% separating set in the MRF over $\{Y, T, \bX^S, \bX_0\}$ does not
% necessarily mean that one does not exist, but if one is found it will
% be sufficient for identifying PATE.

To estimate a MRF, we use a mixed graphical model
\citep{yang2015graphical, haslbeck2020mgm}, which allows for both continuous and categorical variables. More concretely, we assume that each node can be modeled as the exponential family distribution using the remaining variables.
\begin{equation}
  \Pr (G_r \mid G_{-r}) = \mbox{exp} \biggl\{ \alpha_r G_r +
  \sum_{h \neq r} \theta_{r,h} G_r G_h + \varphi(G_r) - \Phi(G_{-r}) \biggr\},
\end{equation}
where $G_{-r}$ is a set of all random variables in a Markov graph except for variable $G_r$, base measure $\varphi(G_r)$ is given by the chosen exponential family, and $\Phi(G_{-r})$ is the normalization constant.
For example, for a Bernoulli distribution, the conditional distribution can be seen as a logistic regression model.
\begin{equation}
  \Pr (G_r \mid G_{-r}) = \cfrac{\mbox{exp}(\alpha_r + \sum_{h \neq r} \theta_{r,h} G_h)}{\mbox{exp}(\alpha_r + \sum_{h \neq r} \theta_{r,h} G_h) + 1}.
\end{equation}
In general, we model each node using a generalized linear model conditional
on the remaining variables. Using this setup, we can estimate the structure of the MRF by estimating parameters $\{\theta_{r,h}\}_{h \neq r}$; $\theta_{r,h} \neq
0$ for variable $G_h$ in the neighbors of variable $G_r$ and
$\theta_{r,h} = 0$ otherwise. We estimate each generalized linear
model with $\ell_1$ penalty to encourage sparsity
\citep{meinshausen2006high}.
Finally, using the AND rule, an edge is estimated to exist between variables $G_r$ and $G_h$ when $\theta_{r,h}
\neq 0$ {\it and} $\theta_{h,r} \neq 0.$ Researchers can also use an alternative OR rule (an edge exists when $\theta_{r,h}
\neq 0$ {\it or} $\theta_{h,r} \neq 0$) and obtain the same theoretical guarantee of graph recovery.

\paragraph{Estimating Separating Sets.}
Given the estimated graphical model, we can enumerate many different
separating sets. First, we focus on the estimation of a separating set
of the smallest size because it often produces more stable weights and
thus improves estimation accuracy.
It is important to note that this separating set might not be the
smallest with respect to the underlying unknown DAG because MRFs don't encode
all conditional independence relationships between variables. It is
the smallest size among all separating sets estimable from MRFs.

We estimate this separating set from pre-treatment covariates $\bZ$ as
an optimization problem. A separating set should block all simple
paths between outcome $Y$ and variables in the sampling set
$\bX^S$. Therefore, we first enumerate all simple paths between $Y$
and $\bX^S$ and then find a minimum set of variables that intersect
all paths.

Define $q$ to denote the number of variables in $\bZ$. We then define
$\bd$ to be a $q$-dimensional decision vector with $d_j$ taking $1$
if we include the $j$ th variable of $\bZ$ into a separating set and taking $0$ otherwise. We use $\bP$ to store all simple paths
from $Y$ to each variable in $\bX^S$ where each row is a $q$-dimensional
vector and its $j$ th element takes $1$ if the path contains the $j$
th variable. With this setup, the estimation of the separating set of the smallest size
is equivalent to the following linear programming problem given the
estimated graphical model.
\begin{eqnarray*}
  && \mbox{ min}_{\bd} \ \sum_{j=1}^q d_j \ \ \ \ \ \mbox{s.t., } \bP \bd \geq \mathbf{1}.
\end{eqnarray*}
where $\mathbf{1}$ is a vector of ones. The constraints above ensure that all simple paths intersect with at least one variable in a selected separating set, and the objective function just counts the total number of variables to be included into a separating set. Therefore, by optimizing this problem, we can find a set of variables with the smallest size that is guaranteed to block all simple paths.

It is important to emphasize that the estimation of the Markov graph
is subject to uncertainty as any other statistical methods. In our application, we incorporate
uncertainties about set estimation through bootstrap.
We also investigate accuracy of the proposed algorithm through
simulation studies in the supplementary material.
We find that estimators based on estimated separating sets often have
similar standard errors to the ones based on the true sampling set. Although our approach introduces an additional estimation step of
finding separating sets to relax data requirements, it does not suffer from substantial efficiency loss.

\paragraph{Incorporating Users' Constraints.}
One advantage of our approach is that we can allow the flexibility for researchers
to explicitly specify variables that they cannot measure in the target
population. %Thus, researchers can estimate a separating set such that
%it can be measured in both the experimental sample and the non-experimental
%target population.
This is important in practice because it is often the case that researchers can measure a large
number of covariates in the experimental data but they can collect relatively few variables
in the target population. We can easily adjust the previous
optimization problem to account for this restriction. Define $\bu$ to be a $q$-dimensional vector with $u_j$ taking $1$
if we want to exclude the $j$ th variable of $\bZ$ from a separating
set and taking $0$ otherwise.  As we define $\bX_0$ to be those variables observed in both the experimental sample and the target population, $\bu$ will place constraints on those covariates in $\bX^S$ that are unobservable.
 Then, the optimization problem above changes as follows.
\begin{eqnarray*}
  && \mbox{ min}_{\bd} \ \sum_{j=1}^q d_j \ \ \ \ \ \mbox{ s.t., } \bP \bd \geq \mathbf{1} \ \mbox{ and } \ \bu^\top \bd =0
\end{eqnarray*}
% Once we have estimated separating sets, we need to assess whether the conditional independence (equation~\eqref{eq:markov}) holds in the experimental data. This exercise is similar to conducting a balance check after matching \citep{ho2007matching}, making sure that our algorithm achieves the desired conditional independence between observed outcomes and a sampling set.

In practice, it is possible that there exists no separating set,
subject to user constraints. In our example, a true separating set
could include social connections, which are not measured in the Northern Uganda
Survey (the population data). In this case, there is no feasible separating
set and our algorithm finds no separating set.

\subsection{Estimation of Population Average Treatment Effect}
\label{subsec:est_pate}
To estimate the PATE with estimated separating sets, we use an inverse probability weighting estimator. First, we
estimate a probability of being in the experiment $\Pr(S_i = 1 \mid
\bW_i),$ for example, using a logistic regression
\citep{Stuart:2011hr, stuart2017trans} with adjustment for the actual
target population size \citep{Buchanan:2018dd}.\footnote{To account
  for the fact that the target population data is a random sample from
  the actual target population, we follow \citet{scott1986fitting,
    Buchanan:2018dd} to estimate a weighted logistic regression. We use
  weights 1 for the experimental data, and use weights $m/(N-n)$ for
  the target population data where the size of the actual target population $N
  = 10,000,000$ (population size in Northern Uganda), the size of the experimental data $n =
  2,598$, and the size of the target population data $m = 21,348$.} Following \citet{Buchanan:2018dd}, we stack
the experimental data and the population data, and $S_i = 1$ ($S_i =
0$) indicates that unit $i$ belongs to the experimental data (the
population data). We can then estimate generalization weights as
\begin{eqnarray}
  \pi_i = \frac{1}{\Pr(S_i = 1 \mid \bW_i)} \times \frac{\Pr(S_i=0 \mid \bW_i)}{\Pr(S_i=0)},
\end{eqnarray}
where a usual inverse probability is adjusted by $\Pr(S_i=0|\bW_i)/\Pr(S_i=0)$ because the PATE is defined only with the
population data, i.e., $\E[Y_i(1) - Y_i(0)|S_i = 0]$.
Finally, we compute the inverse probability weighting estimator \citep{Stuart:2011hr}.
\begin{eqnarray}
  \hat{\tau} \equiv
  \cfrac{\sum_{i; S_i= 1} \pi_i p_i T_iY_i}{\sum_{i; S_i= 1} \pi_i p_i
  T_i}  -   \cfrac{\sum_{i; S_i= 1} \pi_i (1-p_i) (1-T_i)Y_i}{\sum_{i;
  S_i= 1} \pi_i (1-p_i) (1-T_i)}, \label{eq:ipw}
\end{eqnarray}
where $p_i \equiv \Pr(T_i = 1 \mid S_i= 1, \bW_i)$ is known by the experimental design.
We prove its consistency in the supplementary material.

Researchers can also employ an outcome-model-based estimator and a doubly
robust estimator for the PATE \citep{hernan2019, dahabreh2020extending}. To maintain the
clear comparison with the original analysis that uses a weighting
approach, we will focus on the inverse probability weighting estimator
(equation~\eqref{eq:ipw}) in Section~\ref{sec:app_ana}, while we also
report results from the other two estimators in the supplementary material (\ref{sec:add-result}).

\subsection{Relationship with Sensitivity Analysis}
\label{subsec:sa}
Here, we want to clarify relationships between our proposed approach
and sensitivity analyses for generalization
\citep[e.g.,][]{Nguyen:2017jw, nguyen2018, andrews2019}. In particular, our
proposed approach can also be used to simplify sensitivity
analysis. Sensitivity analysis in general uses some sensitivity
parameters to quantify certain aspects of unobserved covariates. For
example, \citet{Nguyen:2017jw, nguyen2018} propose a sensitivity parameter that
captures the distribution of covariates unmeasured in the target population, and \citet{andrews2019} introduce a sensitivity parameter
to quantify the predictive power of unobserved covariates relative to
that of observed covariates. When there is a much richer set of covariates in the
experimental data than the target population data --- the common scenario and the
main focus of our paper, analysts naturally need to specify a large
number of sensitivity parameters to account for such potentially many
covariates that are not measured in the target population data.

In such scenarios, it is well known that  sensitivity parameters become more difficult to interpret, and analysts
need to add more parametric assumptions (e.g., additivity) to handle many
unobserved variables. Our proposed approach can be used
to first find the smallest separating set within the experimental
data. Then, researchers do not need to consider all covariates that
are unmeasured in the target population data, and they can focus on a
potentially much smaller set of covariates estimated as a separating
set. Researchers can then use their preferred sensitivity analysis
technique to deal with a much smaller set of covariates that are unmeasured in
the target population data.

It is also important to clarify the scope of our approach. While we
relax a stringent assumption that a sampling set is measured in both
experimental and target population data, Theorem~\ref{thm_id_sep} still assumes that a
sampling set is observed at least in the experimental data. If
researchers are worried that a sampling set is not observed even in
the experimental data (i.e., not observed in either the experimental or
target population data), sensitivity analysis for completely
unobserved covariates might be of greater importance.

\section{Empirical Analysis}
\label{sec:app_ana}
Applying the proposed method, we examine the YOP described in
Section~\ref{sec:app}. Our focus is on a central methodological challenge of
covariate selection. In the original analysis, the authors adjusted
for all eight variables shared by the experimental and population data.
However, as noted in the original paper, it is unknown whether the
original eight variables is a separating set necessary for estimating PATEs. To
tackle this pervasive concern, we employ the proposed approach and
select a separating set under two different
assumptions about a sampling set and a heterogeneity set.

First, we incorporate domain knowledge about a heterogeneity set,
while we maintain the original assumption about a sampling set.
As explained in Section~\ref{sec:sep}, by combining substantive information about a sampling set and a heterogeneity set,
we can find a separating set, which can be much smaller than each one
of the two. Relying on this smaller separating set, we find that point
estimates are similar to estimates based on the original sampling set,
but standard errors of the
proposed approach are smaller for 14 out of 17
outcomes that the original analysis studied. Incorporating domain
knowledge about a heterogeneity set can help us find a smaller set of
variables sufficient for the PATE estimation, thereby improving efficiency.

Second, we relax the original assumption about a sampling set --- the shared
eight variables contain all relevant variables, and we allow for two
additional unobserved variables. In the conventional approach based
on a sampling set, researchers cannot estimate PATEs under this
assumption. In contrast, the proposed approach estimated appropriate
separating sets for 12 out of 17 outcomes, and thus, we can estimate the PATE
for those 12 outcomes even with the two additional
unobserved sampling variables. At the same time, we reveal that
estimated PATEs are sensitive to the original assumption about the
sampling mechanism for the other 5 outcomes.

The original experiment used clustered randomization, and therefore,
we compute clustered standard errors at the group level, which is the
level at which treatments were randomly assigned. While we maintain
the no interference assumption \citep{Rubin:1990tm} we made in Section~\ref{subsec:setup}, which is the standard assumption
in the generalization literature, if researchers wish to account for
interference within groups, it is important to additionally
account for the difference in group structure between the experimental
and target population data. Future work is necessary to consider this intersection
of interference and generalization literature.

\subsection{Incorporating Domain Knowledge on Heterogeneity Set}
\label{subsec:exact-app}
To begin with, we maintain an assumption about a sampling set in the original
analysis, i.e., $\bX^S =$ \{\texttt{Gender, Age, Urban, Marital status, School
  attainment, \\ Household size, Durable assets, District}\}.
Although the original analysis relies only on this knowledge of the sampling set for the PATE
estimation, the authors also carefully discuss a heterogeneity set in their
paper. In particular, they discuss two variables: gender and initial
credit constraints. There are two natural covariates in the
experimental data that capture these concepts, \texttt{Gender} and
\texttt{Initial Saving}, respectively. Importantly, however,
\texttt{Initial Saving} is measured only in the experimental sample
and not in the target population data. Thus, $\bX^H =$
\{\texttt{Gender, \fbox{Initial Saving}}\} where the square box
represents a variable unmeasured in the target population.

In existing approaches, when a subset of heterogeneity variables are unmeasured in
the target population as in this case, it is difficult to incorporate such domain knowledge
into analysis and researchers often ignore the heterogeneity
set altogether. In contrast, our proposed method uses knowledge about the heterogeneity set
to estimate an exact separating set, which is potentially smaller than the observed
sampling set and thus can increase the estimation accuracy for the PATE
estimation. We first estimate a Markov random field over the
union of sampling and heterogeneity sets within the experimental
data. Then, we select an exact separating set that makes sampling set $\bX^S$ and heterogeneity set
$\bX^H$ conditionally independent under a constraint that \texttt{Initial Saving} is unmeasured in the population data and
cannot be selected. To take into account uncertainties, we estimate
Markov random fields and select exact separating sets in each of 1000 bootstrap samples.

\begin{figure}[!t]
  \begin{center}
    \begin{tabular}{c}
      % 1
      \begin{minipage}{0.65\hsize}
        \begin{center}
          \includegraphics[height = 2in]{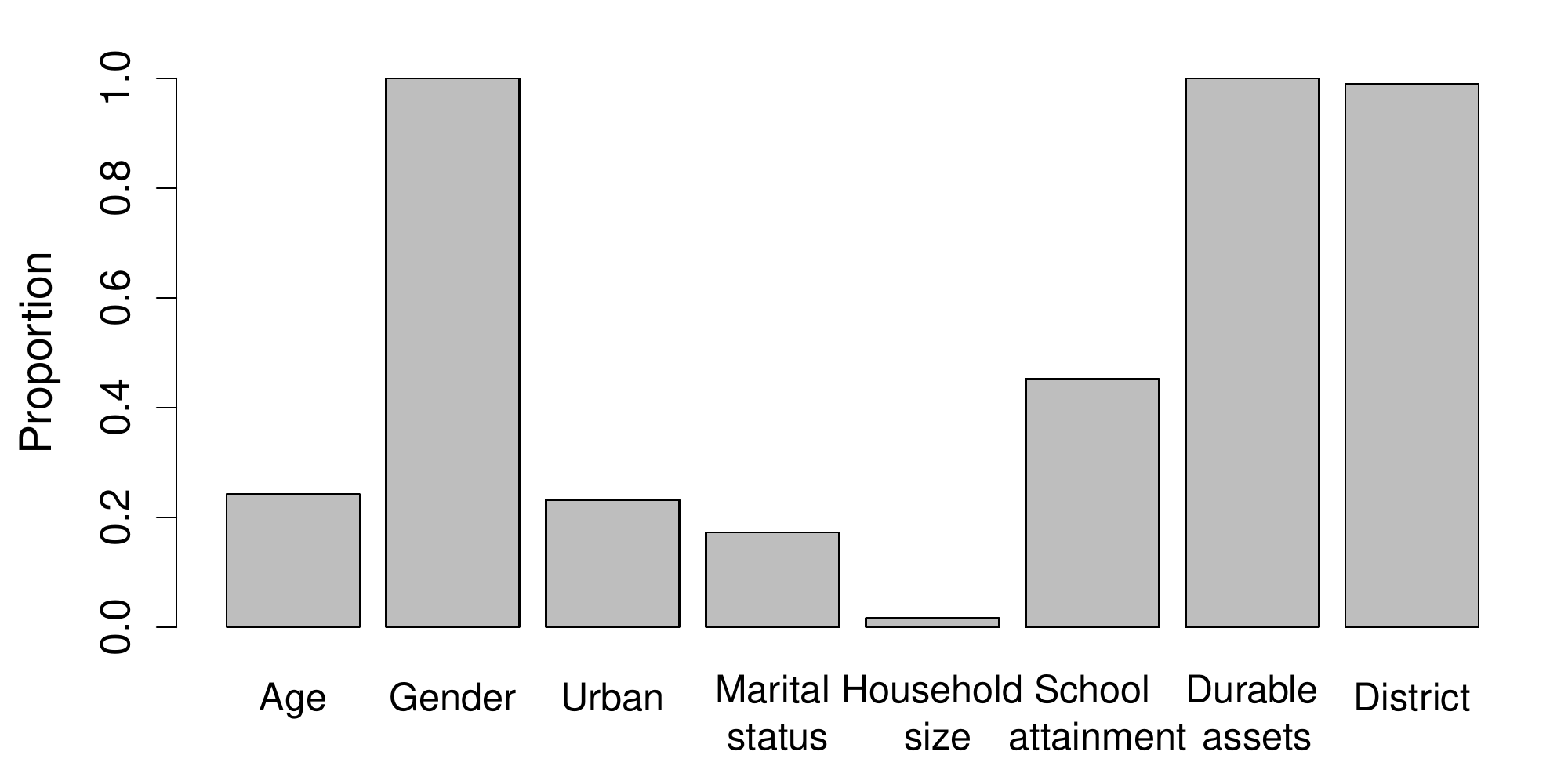}
          (a) Proportion of Being Estimated \\[-5pt] as Separating
          Sets.
        \end{center}
      \end{minipage}
      \hspace{-0.2in}
      % 2
      \begin{minipage}{0.3\hsize}
        \begin{center}
          \includegraphics[height  = 2in]{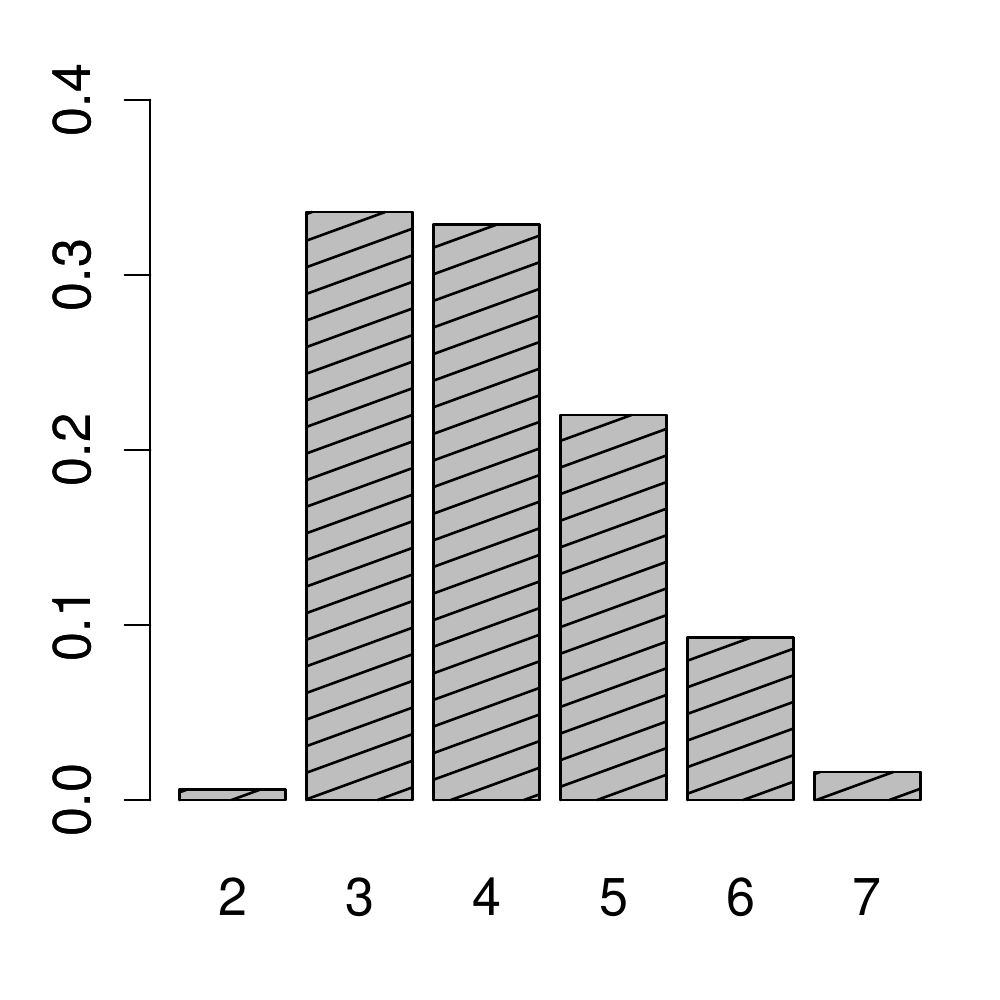}
          (b) Size of Estimated \\[-5pt] \ \ \ Separating Sets.
        \end{center}
      \end{minipage}
    \end{tabular}
    \spacingset{1.2}{\caption{Estimated Exact Separating Sets. {\it
          Note:} Panel (a) shows the proportion of each variable being
        estimated to be in an exact separating set over 1000 bootstrap
        samples. Panel (b) reports
        the size of estimated exact separating sets over 1000
        bootstrap samples.}\label{fig:cov_XH}}
  \end{center}
\end{figure}

Figure~\ref{fig:cov_XH} reports the results. The left panel (a) shows
the proportion of each variable being estimated to be in an exact
separating set over 1000 bootstrap samples. As the definition of
separating sets (Definition~\ref{sep}) implies, the intersection of
sampling and heterogeneity set, \texttt{Gender}, is always
selected. In addition, \texttt{Durable assets} and \texttt{District} are
selected almost always. Importantly, when we look at the size of
estimated exact separating sets (the right panel (b)), it is often
much smaller than the original size of eight (the mean size is
$4.11$). This means that even when a sampling set is sufficient for
estimating the PATE, researchers can find a smaller separating set by
incorporating domain knowledge of heterogeneity sets with the proposed
approach.

If assumptions about the sampling set and the heterogeneity set hold, estimators based on the original sampling set
$\bX^S$ and on the estimated separating sets $\bW$ are
both consistent. However, standard errors of the latter might be
smaller because corresponding estimated weights might be more stable.

To estimate the PATEs, we use the inverse probability weighting
estimator proposed in Section~\ref{subsec:est_pate}. First, we
estimate weights using the following logistic regression.
\begin{eqnarray}
  \mbox{logit}\{\Pr(S_i = 1 \mid \bC_i)\} = \alpha_0 + \bC_i^\top \beta,
\end{eqnarray}
where $\bC = \bX^S$ for the estimator based on the original sampling
set and $\bC = \bW$ for our
proposed estimator. We stack the experimental data (sample size =
$2,598$) and the population data (sample
size = $21,348$) and $S_i = 1$ ($S_i = 0$) indicates that unit $i$ belongs
to the experimental data (the population data). We can then estimate
weights as  $\hat\pi_i = 1/\widehat{\Pr}(S_i = 1 \mid \bC_i)
\times \widehat{\Pr}(S_i=0 \mid \bC_i)/\widehat{\Pr}(S_i=0),$
as proposed in Section~\ref{subsec:est_pate}. Note that treatment
assignment probability in the experiment $\Pr(T_i =1 \mid S_i = 1,
\bW_i)$ is equal to $\Pr(T_i =1 \mid S_i = 1, \mathbf{D}_i)$ where
$\mathbf{D}_i$ is a vector indicating 14 districts,
because the treatment randomization was stratified by  districts
\citep{blattman2013generating}.
We use the block bootstrap to compute standard errors clustered at the
group level as done in the
original analysis. To take into account uncertainties
of both steps --- the estimation of separating sets and that of the
PATE, in each of 1000 bootstrap samples, we estimate separating sets,
estimate weights, and then estimate the PATE. Note that the difference between the estimator
based on the original sampling set and our proposed estimator comes
only from the selection of covariates
$\bC$ in the estimation of weights.

\begin{table}[!t] \centering
  \small
  \resizebox{\textwidth}{!}{
    \begin{tabular}{@{\extracolsep{5pt}} L{2.1in}C{0.6in}C{0.6in}|L{2in}C{0.6in}C{0.6in}}
      \hline  \hline
      & Original & Sep. Set &  & Original & Sep. Set  \\[-5pt]
      & estimate & estimate &  & estimate & estimate \\ \hline
      \normalsize \bf \underline {Employment} &  &  &  \normalsize \bf \underline {Investments} &  &  \\
      Average employment hours & 5.87 & 4.79 & Vocational training & 0.51 & 0.53 \\[-3pt]
      & (3.25) & (2.39) &  & (0.07) & (0.05) \\
      Agricultural & 0.75 & 0.30 & Hours of vocational training & 277.18 & 337.59 \\[-3pt]
      & (2.01) & (1.69) &  & (51.21) & (40.77) \\
      Nonagricultural & 5.12 & 4.49 & Business assets & 400.31 & 425.02 \\[-3pt]
      & (2.62) & (1.79) &  & (125.19) & (135.65) \\
      Skilled trades only & 2.73 & 4.36 &  &  &  \\[-3pt]
      & (1.56) & (0.99) & \normalsize \bf \underline {Business Formality} &  &  \\
      No employment hours & 0.00 & -0.03 & Maintain records & 0.19 & 0.20 \\[-3pt]
      & (0.02) & (0.03) &  & (0.07) & (0.07) \\
      Any skilled trade & 0.24 & 0.27 & Registered & 0.07 & 0.09 \\[-3pt]
      & (0.07) & (0.06) &  & (0.04) & (0.05) \\
      Works mostly in a skilled trade & -0.01 & 0.04 & Pays taxes & -0.01 & 0.05 \\[-3pt]
      & (0.03) & (0.03) &  & (0.06) & (0.05) \\
      \normalsize \bf \underline {Income} &  &  & \normalsize \bf \underline {Urbanization} &  &  \\
      Cash earnings & 12.15 & 12.54 & Changed parish & -0.03 & -0.01 \\[-3pt]
      & (5.3) & (5.11) &  & (0.07) & (0.04) \\
      Durable assets & 0.09 & 0.18 & Lives in Urban area & -0.06 & -0.01 \\[-3pt]
      & (0.16) & (0.13) &  & (0.07) & (0.04) \\
      \hline
    \end{tabular}}
  \spacingset{1.2}{
    \caption{Estimates of Population Average Treatment Effects based on
      the Original Set and the Estimated Separating Set. {\it Note:} We estimated
      population average treatment effects of 17 outcomes
      using weights based on the original eight variables (``Original
      estimate'') and the estimated exact separating set (``Sep. Set
      estimate''). Standard errors of the proposed estimators are smaller for 14 out of 17 outcomes.}\label{tab:estimate_XH}}
\end{table}

We report results in Table~\ref{tab:estimate_XH}. Effects of the YOP
are large and positive across many outcomes even among the broader
target population. For example, the average employment hours would
increase by $4.79$ hours (19\% increase compared to the control group), monthly
cash earnings would increase by $12,540$ Uganda shilling (36\% increase),
and a proportion of people enrolled in vocational training would increase
by $53$ percentage points (349\% increase). Comparing estimates based on
the original sampling set and those based on the proposed separating set, we
reveal that point estimates are similar to estimates with the
original eight variables, and differences between them are not statistically significant at
the conventional 0.05 level. This is expected because both estimators
are consistent under the assumption that both specified sampling and
heterogeneity sets are correct. More interestingly, we find that, for
14 out of 17 outcomes, standard errors of estimators based on the estimated separating sets
are smaller than those based on the original sampling set. On
average, standard errors of the proposed approach are about 16\%
smaller. For the outcome ``Lives in Urban area,'' the standard error
reduces more than 45\%.   This shows that by incorporating domain knowledge about heterogeneity sets, we
can estimate smaller separating sets, which often improve efficiency.

% \begin{figure}[!t]
% \hspace{-0.25in}
%     \includegraphics[width = 1.05\textwidth]{figs/Se_XH_final_0413.pdf}
%     \vspace{-0.5in}
%     \spacingset{1.2}{\caption{Comparing Standard Errors of Population
%         Average Treatment Effects Estimates based on the Original Set and the Proposed Separating Set.}\label{fig:se_XH}}
% \end{figure}

\subsection{Accounting for Unobserved Sampling Set}
\label{subsec:mar-app}
In the previous analysis, we maintained the original authors' assumption about the
sampling set and additionally took into account the assumption about
the heterogeneity set. Here, we focus on estimating PATEs under weaker
assumptions and directly address a concern noted in
the original paper that the shared eight variable might not contain
all relevant variables. In particular, \citet{blattman2013generating}
discuss two potentially problematic variables. First, the authors are concerned that
when the government screened applications at the village level, people
with more social connections may have received some privilege. Second, people with ``affinity for
entrepreneurship'' \citep{blattman2013generating} might have been more
likely to apply for the program in the first place. To account for
these two sources of sample selection, we assume that a
true sampling set contains two additional variables: (1)
\texttt{Connection}, the number of community groups that a respondent
belongs to, as a measure of social connections, and (2)
\texttt{Business Advice}, total hours spent getting business advice in last 7 days, as a measure of initial motivation and affinity
for entrepreneurship. Importantly, both of these two variables are not measured in the population data. Therefore, $\bX^S =
$ \{\texttt{Gender, Age, Urban, Marital status, School attainment,
  Household size, Durable assets, District, \fbox{Connection},
  \fbox{Business Advice}}\} where the last two variables are measured only in
the experiment and not in the population data. Moreover, we don't
make any assumption about heterogeneity sets. Under this assumption,
the current practice based on sampling sets or heterogeneity sets cannot estimate any PATEs;
weights can be estimated only when sampling sets or heterogeneity sets
are measured in both the experimental and population data. In
contrast, the proposed method can select appropriate separating sets,
should they exist, under such data constraints.

% \begin{figure}[!t]
% \hspace{-0.25in}
%     \includegraphics[width = 1.05\textwidth]{figs/Prop_nofeasible_large_0413.pdf}
%     \spacingset{1}{\caption{Proportions of Infeasible Solutions. {\it
%           Note:} We estimated marginal separating sets under a constraint
%         that two sampling variables are unobserved in the population
%         data. For four outcomes, proportions of infeasible solutions are below
%         30\%.}\label{fig:feasible_XS}}
% \end{figure}

\begin{figure}[!t]
\hspace{-0.25in}
    \includegraphics[width = 1.05\textwidth]{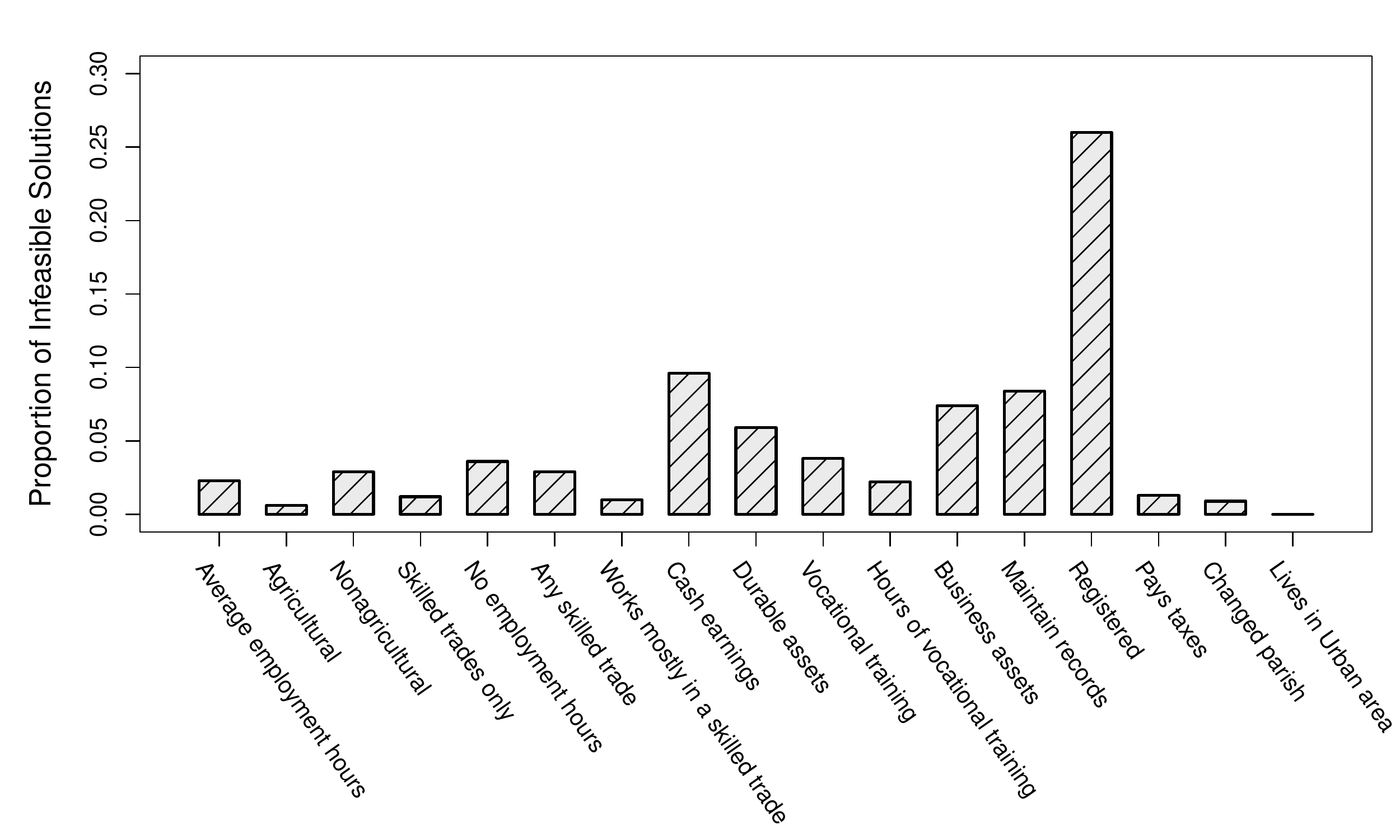}
    \spacingset{1}{\caption{Proportions of Infeasible Solutions. {\it
          Note:} For 17 outcomes, we estimated marginal separating sets under a constraint
        that two sampling variables are unobserved in the population
        data. The figure shows the proportion of infeasible solutions for each outcome.}\label{fig:feasible_XS}}
\end{figure}

There are two questions of interest for each outcome; (1) Can we find a separating set and estimate the PATE?
(2) If we can estimate the PATE, is an estimate different from the
one based on the original eight variables? We estimate marginal separating sets using the proposed algorithm. For
each outcome $Y$, we first estimate a Markov random field and then
select a separating set that makes outcome $Y$ and sampling set
$\bX^S$ conditionally independent under a constraint that the two
unobserved variables (\texttt{Connection, Business Advice}) cannot be
selected. When the algorithm can find no separating set under the
constraint, we call it an ``infeasible solution.'' We use
block bootstrap to compute standard errors clustered at the group
level as done in the original analysis. To take into account uncertainties over the
covariate selection, we estimate Markov random fields and select
separating sets in each of 1000 bootstrap samples.

We begin by computing proportions of infeasible solutions among the 1000 bootstraps
(Figure~\ref{fig:feasible_XS}). Proportions vary across outcomes,
ranging from 0.0\% (``Lives in Urban area'') to 26.0\% (``Registered''), and on average, 4.70\%.  Given that the current practice
just based on sampling or heterogeneity sets cannot estimate PATEs for
any outcomes, it is interesting that the proportions of infeasible
solutions are smaller than 5\% for 12 out of 17 outcomes. For the
remaining five outcomes, the average proportion of infeasible
solutions is 11.5\%, suggesting that the PATE estimates for these
outcomes are sensitive to unobserved sampling variables,
\texttt{Connection} and \texttt{Business Advice}.

\begin{figure}[!t]
    \centering
    \includegraphics[width = 6.0in]{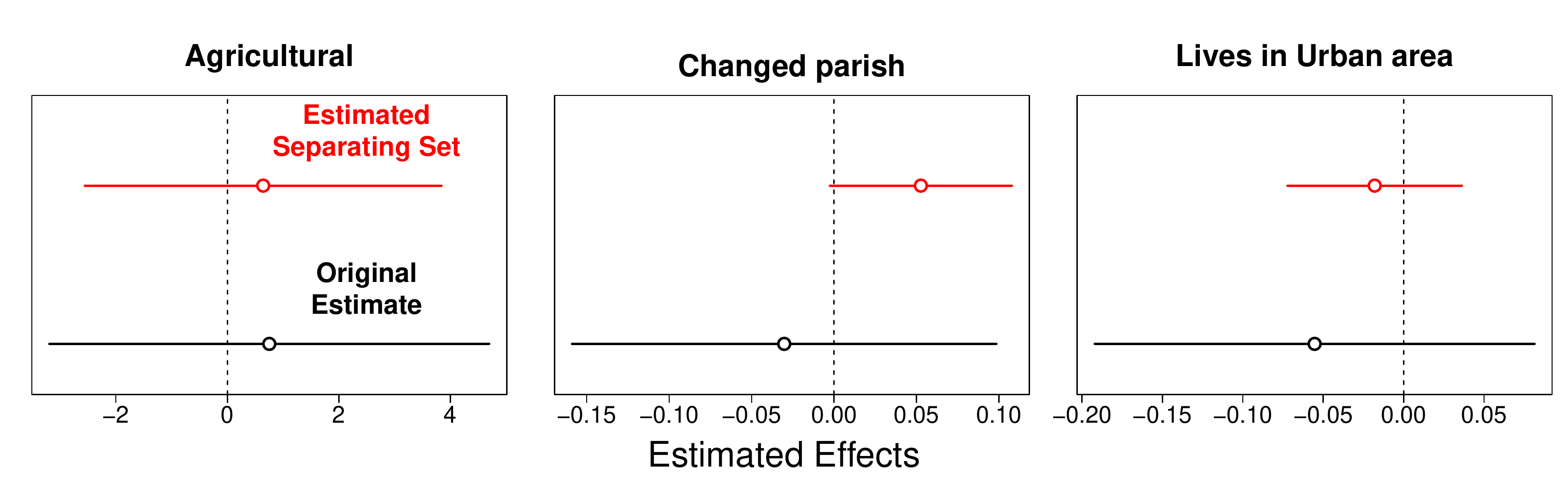}
    \spacingset{1}{\caption{Estimates of Population Average Treatment Effects based on
        the Original Sets and Estimated Marginal Separating Sets. {\it Note:} We estimated
        population average treatment effects for 3 outcomes that have
        estimated proportions of infeasible solutions below
        1\%. Weights are based on the original eight variables
        (``Original'') and estimated marginal separating sets (``Estimated
        Separating Set''). }\label{fig:estimate_XS}}
\end{figure}

For three outcomes that have less than 1\% of infeasible
solutions,\footnote{\spacingset{1}{\footnotesize ``Agricultural''[0.6\%], ``Changed parish'' [0.9\%],  ``Lives in Urban area''[0.0\%].}} we also report estimates with 95\% confidence
intervals in Figure~\ref{fig:estimate_XS}. We use the block bootstrap to
compute standard errors clustered at the group level. To take into account
uncertainties of both steps --- the estimation of separating sets and
that of the PATE, in each of 1000 bootstrap samples, we estimate
separating sets, estimate weights, and then estimate the PATE. We find that point estimates are similar
to the original estimates for the two outcomes (``Agricultural'' and
``Lives in Urban area''), and for ``Changed parish,'' while point
estimates are slightly different due to large standard errors, the
difference between the original estimate and the estimate based on an
estimated separating set is not statistically significant at the
conventional level of $0.05.$ For all three outcomes, standard errors are smaller than
the original ones. That is, estimates of the PATEs are robust to alternative separating sets, i.e., even if the
sampling set includes additional unobserved variables, substantive
conclusions are similar. This result demonstrates that the proposed
algorithm of selecting separating sets allows researchers to estimate
PATEs in situations where previous methods could not.

% \begin{figure}[!t]
%     \centering
%     \includegraphics[width = \textwidth]{figs/Estimate_XS_large_0413.pdf}
%     \spacingset{1}{\caption{Estimates of Population Average Treatment Effects based on
%         the Original Sets and Estimated Marginal Separating Sets. {\it Note:} We estimated
%         population average treatment effects for 4 outcomes that have
%         estimated proportions of infeasible solutions below
%         30\%. Weights are based on the original eight variables
%         (``Original'') and estimated marginal separating sets (``Estimated
%         Separating Set''). }\label{fig:estimate_XS}}
% \end{figure}

\section{Concluding Remarks}
The increased emphasis on well-identified causal effects in the social
and biomedical sciences can sometimes lead researchers to narrow the
focus of their research question and limit their findings to the
experimental sample.  However, primary research questions are often
driven by the need to discover the impact of an intervention on a
broader population.  The extant literature has focused on the
mathematical underpinnings concerning the generalizability of
experimental evidence.  The aim of this paper is to provide applied
researchers with a means for uncovering a separating set using the experimental data alone.

Building on previous approaches, we clarify the role
of the separating set -- and its relationship to the sampling
mechanism and treatment effect heterogeneity -- in identification of
population average treatment effects.  It makes clear that
there are many possible covariate sets researchers can use for the
recovery of population effects, and it allows us to develop a new
algorithm that can incorporate researchers' data constraints on the
target population.

As a concrete context, we focus on the YOP in Uganda. For these types of large-scale development programs,
potential benefits and necessity of generalization are well known
among researchers and policy makers. However, analysts are often
constrained by available covariate information, which limits applicability
of existing approaches that assume rich covariate data from both the
experimental and population samples. Our proposed algorithm can help
researchers to estimate appropriate separating sets, if any should exist, even under
such data constraints. We find that by incorporating domain knowledge
about heterogeneity sets, which is often overlooked in the PATE
estimation, we can substantially improve efficiency. We also reveal
that the proposed algorithm can find separating sets for 12 out of 17
outcomes, even if we allow for two additional sampling variables that
are not measured in the population.

Identifying population effects remains a challenging task for
experimental researchers.  The results here suggest researchers can increase a chance of
generalization by collecting rich covariate information on their
experimental subjects, even when their capacity of the population data collection is limited.

\vspace{0.3in}
\spacingset{1.4}
{\small
\pdfbookmark[1]{References}{References}
\printbibliography}

\clearpage
\appendix

\begin{refsection}

\begin{center}
  {\bf \LARGE Supplementary Material} \vspace{0.2in}\\
  {\large Covariate Selection for Generalizing Experimental
      Results}
\end{center}

\renewcommand{\thefigure}{SM-\arabic{figure}}
\renewcommand{\thetable}{SM-\arabic{table}}
\renewcommand\thesection{SM-\arabic{section}}
\renewcommand\thesubsection{\thesection.\arabic{subsection}}

\section{Proof of Theorems}\label{app:proofs}
Here, we provide proofs for the theorems presented in the paper.
% \subsection{Proof of Theorem~\ref{pate}}
% \label{subsec:pate}\vspace{-0.1in}
% \begin{eqnarray*}
%     \tau & = & \int_w \E[Y(1) - Y(0) \mid S=0, \bW=\bw] d F_{\bW \mid S = 0} (\bw) \\
%          & = & \int_w \E[Y(1) - Y(0) \mid S =1, \bW=\bw] d F_{\bW \mid S = 0} (\bw) \\
%          & = & \int_w \biggl\{\E[Y(1) \mid S=1, \bW=\bw] - \E[Y(0) \mid S=1, \bW=\bw]\biggr\} d F_{\bW \mid S = 0} (\bw) \\
%          & = & \int_w \biggl\{\E[Y(1) \mid T =1, S =1, \bW=\bw] - \E[Y(0) \mid T =0, S=1, \bW=\bw]\biggr\} d F_{\bW \mid S = 0} (\bw) \\
%          & = & \int_w \biggl\{\E[Y \mid T =1, S =1, \bW=\bw] - \E[Y \mid T =0, S =1, \bW=\bw]\biggr\} d F_{\bW \mid S = 0} (\bw).
%   \end{eqnarray*}
% where the first equality follows from the rule of the conditional
% expectation, the second from the definition of a separating set
% (Definition~\ref{sep}), the third from the
% linearity of the expectation, the fourth from Assumption~\ref{random},
% and the final follows from the consistency condition. \qed

\subsection{Proof of Theorem~\ref{thm_id_Exsep}}
\label{subsec:Exsep}
In this proof, we assume that the separating set $\bW$ is disjoint with the sampling set $\bX^S$ and the heterogeneity set $\bX^H$ for simpler notations. The same proof applies to the case in which some variables of the sampling set or the heterogeneity set are in the separating set. First, we have
\begin{eqnarray}
  \bX^H \ \indep \ \bX^S \mid \bW, T, S=1. \label{eq:est1}
\end{eqnarray}
From Random Treatment Assignment(Assumption~\ref{random}), we have
\begin{eqnarray}
  T \ \indep \ \bX^S \mid \bW, S=1. \label{eq:est2}
\end{eqnarray}
Combining equations~\eqref{eq:est1} and~\eqref{eq:est2} (Contraction in \citet{pearl2000causality}),
\begin{eqnarray}
  && \{\bX^H, T\} \ \indep \ \bX^S \mid \bW, S=1, \notag
\end{eqnarray}
which implies $\bX^H \ \indep \ \bX^S \mid \bW, S=1.$ Given that the
conditional independence structure of $(\bX^H, \bX^S, \bW)$ is the
same under $S=1$ and $S=0$ (because $S$ only changes the treatment assignment),  we have
\begin{equation}
    \bX^H \ \indep \ \bX^S \mid \bW, S. \label{eq:est3}
\end{equation}
From the definition of the sampling variable,
\begin{eqnarray}
  \bX^H \ \indep \ S \mid \bW, \bX^S. \label{eq:est4}
\end{eqnarray}
Combining equations~\eqref{eq:est3} and~\eqref{eq:est4} (Intersection \citep{pearl2000causality}), we have
\begin{eqnarray}
  && \bX^H \ \indep \ \{S, \bX^S\} \mid \bW, \notag
\end{eqnarray}
which implies
\begin{eqnarray}
  & & \bX^H \ \indep \ S \mid \bW. \label{eq:42}
\end{eqnarray}
Additionally, based on the definition of the heterogeneity set,
\begin{eqnarray}
  && Y(1) - Y(0) \ \indep \ S  \mid \bW, \bX^H. \label{eq:43}
\end{eqnarray}
Therefore, by combining equations~\eqref{eq:42} and~\eqref{eq:43} based on Contraction in \citet{pearl2000causality},
\begin{eqnarray*}
  && \{Y(1) - Y(0), \bX^H\} \ \indep \ S  \mid \bW,
\end{eqnarray*}
which implies $Y(1) - Y(0) \ \indep \ S \mid \bW.$ \qed

\subsection{Proof of Theorem~\ref{thm_id_sep}}
\label{subsec:sep}
First, we have
\begin{eqnarray}
  Y \ \indep \ \bX^S \mid \bW, T, S=1. \label{eq:est5}
\end{eqnarray}
From Random Treatment Assignment(Assumption~\ref{random}), we have
\begin{eqnarray}
  T \ \indep \ \bX^S \mid \bW, S=1. \label{eq:est6}
\end{eqnarray}
Combining equations~\eqref{eq:est5} and~\eqref{eq:est6} (Contraction
in \citet{pearl2000causality}),
\begin{eqnarray}
  && \{Y, T\} \ \indep \ \bX^S \mid \bW, S=1, \notag
\end{eqnarray}
which implies
\begin{eqnarray}
  & &  Y(t) \ \indep \ \bX^S \mid \bW, S=1.
\end{eqnarray}

Given that the conditional independence structure of $(Y(1), Y(0), \bX^S, \bW)$ is the
same under $S=1$ and $S=0$ (because $S$ only changes the treatment assignment, relationship for
potential outcomes and pre-treatment variables would not change),  we have
\begin{equation}
    Y(t) \ \indep \ \bX^S \mid \bW, S, \label{eq:est7}
\end{equation}
for $t = \{0, 1\}.$

From the definition of the sampling variable, for $t = \{0, 1\},$
\begin{eqnarray}
  Y(t) \ \indep \ S \mid \bW, \bX^S. \label{eq:est8}
\end{eqnarray}
Combining equations~\eqref{eq:est7} and~\eqref{eq:est8} (Intersection
in \citet{pearl2000causality}), we have
\begin{eqnarray*}
  && Y(t) \ \indep \ \{S, \bX^S\} \mid \bW,
\end{eqnarray*}
which implies
\begin{eqnarray*}
  & & Y(t) \ \indep \ S \mid \bW
\end{eqnarray*}
for $t = \{0, 1\}.$ This completes the proof.
\qed

\clearpage
\section{IPW Estimator}
\label{sec:ipw}
Here, we show that $\hat{\tau} \xrightarrow{p} \E[Y_i (1) - Y_i(0) \mid S_i = 0].$
\paragraph{Proof}
First, we rewrite the IPW estimator as follows.
\begin{equation}
  \hat{\tau} =
  \cfrac{\frac{1}{n+m}\sum_{i} S_i \pi_i p_i
    T_iY_i}{\frac{1}{n+m}\sum_{i} S_i\pi_i p_i T_i}  -  \cfrac{\frac{1}{n+m}\sum_{i} S_i\pi_i (1-p_i) (1-T_i)Y_i}{\frac{1}{n+m}\sum_{i} S_i\pi_i (1-p_i) (1-T_i)},
\end{equation}
where $n$ ($m$) is the sample size of the experimental data (the
population data). By the law of large number,
\begin{align*}
  \frac{1}{n+m}\sum_{i} S_i \pi_i p_i T_i \xrightarrow{p} \E[S_i
                                                \pi_i p_i T_i] & = \E_{\bW} \{\pi_i \Pr(S_i=1 \mid \bW_i) p_i  \Pr(T_i = 1\mid S_i=  1, \bW_i)\}\\
  & = \E_{\bW} \l\{\frac{\Pr(S_i=0 \mid \bW_i)}{\Pr(S_i=0)}\r\} = 1.
\end{align*}
Similarly, $ \frac{1}{n+m}\sum_{i} S_i \pi_i (1-p_i) (1-T_i)
\xrightarrow{p} 1.$ Again, by the law of large number,
\begin{align*}
  & \frac{1}{n+m}\sum_{i} S_i \pi_i p_i T_iY_i \xrightarrow{p}
    \E[S_i \pi_i p_i T_i Y_i], \hspace{0.1in} \frac{1}{n+m}\sum_{i} S_i \pi_i (1-p_i) (1-T_i) Y_i \xrightarrow{p} \E[S_i \pi_i (1-p_i) (1-T_i) Y_i].
\end{align*}
Hence, $\hat{\tau} \xrightarrow{p} \E[S_i \pi_i p_i T_i Y_i  -
S_i \pi_i (1-p_i) (1-T_i) Y_i].$ We focus on the term on the right.
{\small
\begin{eqnarray*}
  \hspace{-0.3in}&& \E \biggl\{\pi_i \bigg(S_i p_i T_i Y_i - S_i (1-p_i) (1 - T_i) Y_i\bigg)\biggr\} = \E_{\bW} \Biggl\{\pi_i\E \biggl\{S_i p_i T_i
                    Y_i - S_i (1-p_i) (1 - T_i) Y_i \mid \bW_i\biggr\} \Biggr\}\\
  \hspace{-0.3in}& = & \E \Biggl\{\pi_i \Pr (S_i = 1 \mid \bW_i) \E
                       \biggl\{ p_i T_i Y_i - (1-p_i)(1 - T_i) Y_i \mid S_i = 1, \bW_i \biggr\}\Biggr\}\\
  \hspace{-0.3in}& = & \E \Biggl\{\pi_i \Pr (S_i = 1 \mid \bW_i)
       \{p_i \E[T_i Y_i \mid S_i = 1, \bW_i] - (1-p_i) \E[(1 - T_i) Y_i \mid S_i = 1, \bW_i]\}\Biggr\}\\
  \hspace{-0.3in}& = & \E \Biggl\{\pi_i \Pr (S_i = 1 \mid \bW_i)
      \bigg(\E[Y_i(1) \mid S_i = 1, \bW_i] - \E[Y_i(0) \mid S_i = 1, \bW_i]\bigg) \Biggr\}\\
  \hspace{-0.3in}& = & \E \Biggl\{\pi_i \Pr (S_i = 1 \mid \bW_i)\E[Y_i(1) -
      Y_i(0) \mid S_i = 1, \bW_i]\Biggr\} =  \E \Biggl\{\pi_i \Pr (S_i = 1 \mid \bW_i)\E[Y_i(1) - Y_i(0) \mid S_i = 0, \bW_i]\Biggr\}\\
  \hspace{-0.3in}& = & \E \Biggl\{\frac{\Pr (S_i = 0 \mid \bW_i)}{\Pr (S_i = 0)} \E[Y_i(1) - Y_i(0) \mid S_i = 0, \bW_i]\Biggr\}\\
  \hspace{-0.3in}& = & \int_{\bW} \Biggl\{\frac{\Pr (S_i = 0 \mid \bW_i)}{\Pr (S_i =
        0)} \E[Y_i(1) - Y_i(0) \mid S_i = 0, \bW_i]\Biggr\} p(\bW) d\bW\\
  \hspace{-0.3in}& = & \int_{\bW} \E[Y_i(1) - Y_i(0) \mid S_i = 0, \bW_i] p(\bW \mid
      S_i = 0) d\bW =  \E[Y_i(1) - Y_i(0) \mid S_i = 0],
\end{eqnarray*}}
\noindent where the first equality follows from the law of conditional expectation
given $\bW$, the second from the conditional expectation given
$S$, the third from the linearity of expectation, the fourth from
the conditional expectation given $T$, the
fifth from the linearity of expectation, the sixth from the definition
of separating $\bW$, the seventh from the definition of $\pi,$ the eight from the rule of expectation, the ninth from
Bayes rule, and the tenth from the rule of expectation.

\clearpage
\section{Markov Random Fields: Review}
\label{sec:mrf-si}
A Markov random field (MRF), also known as an undirected graphical model, is
a popular statistical model that encodes the conditional independence
structure over multiple observed random variables. The main advantage
of the MRF is that it encodes the conditional independence relationships
of many random variables compactly. While many
important results have been derived for MRFs, we focus on one key
property, so-called, the global Markov property, which we use in our paper.

\begin{figure}[!h]
  \begin{center}
    \includegraphics[width = 0.4\textwidth]{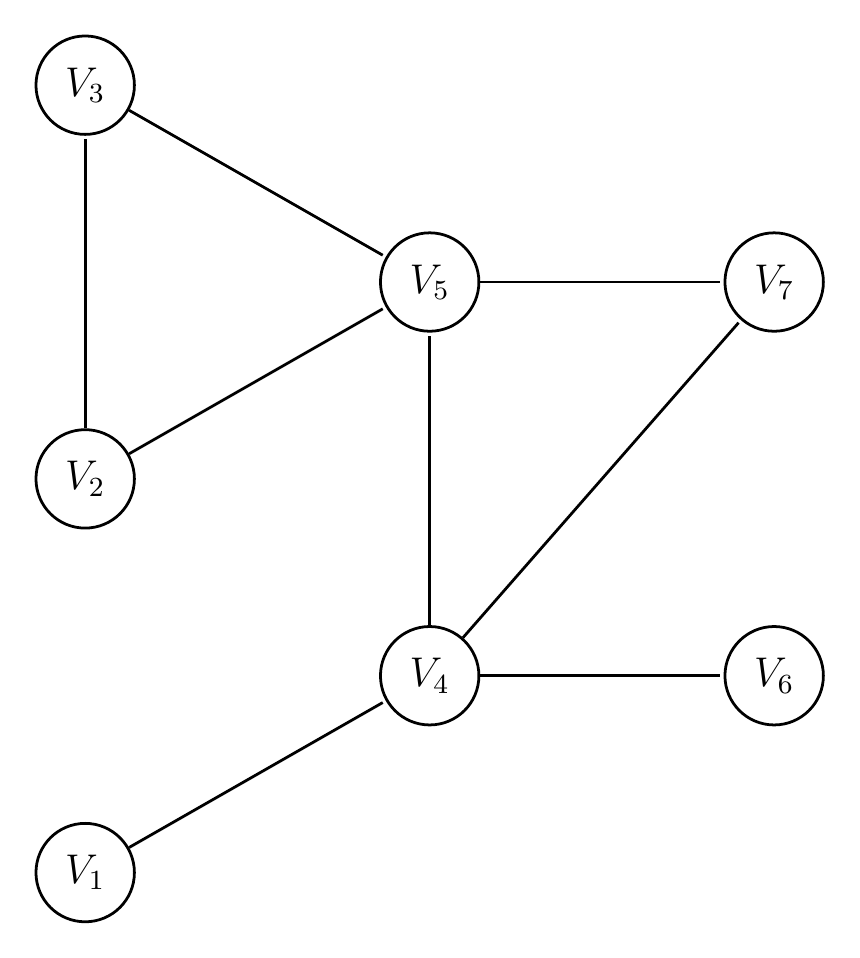}
    \spacingset{1.2}{\caption{Example of a Markov Random Field (MRF).}\label{fig:mrf}}
  \end{center}
\end{figure}

MRFs define the conditional independence relationships via simple
graph separation rules \citep{lauritzen1996graphical}. For sets of
nodes $A$, $B$, and $C$, $A \ \indep \ B \mid C$ if and only if there is no path connecting $A$ and $B$ when
nodes in $C$ are removed from the graph (i.e., nodes in $C$ separates
nodes $A$ and $B$). For example, in Figure~\ref{fig:mrf}, suppose $A =
\{V_1, V_2, V_3\}$ and $B = \{V_6, V_7\}$. Then, if we define $C =
\{V_4, V_5\}$, there is no path connecting $A$ and $B$ once nodes in
$C$ are removed from the graph. Therefore, Figure~\ref{fig:mrf}
encodes the conditional independence relationship, $\{V_1, V_2, V_3\} \ \indep \ \{V_6,
V_7\} \mid V_4, V_5$.

As emphasized in the paper, we use the MRF as the statistical model to
characterize the conditional independence relationships between
observed random variables. We do not use the MRF as a step to estimate
the underlying causal DAG.

\clearpage
\section{Additional Results on Empirical Analysis}
\label{sec:add-result}
In Section~\ref{sec:app_ana},  we focused on the inverse probability weighting estimator
(equation~\eqref{eq:ipw}) to maintain the clear comparison with the
original analysis that uses the weighting approach. In this section,
we report results based on an outcome-model-based estimator\footnote{
  For outcome-model-based estimators, it is unclear whether adjusting for a smaller set of
  covariates leads to an increase in estimation efficiency; it will
  depend on how predictive are those covariates. However, at least in
  our application, we see below in Table~\ref{tab:tab-exact-out} that
  outcome-model-based estimators based on estimated separating sets
  have smaller standard errors than those based on the original
  sampling set for 16 out of 17 outcomes. For
  outcome-model-based estimators, another benefit of having a smaller valid
  separating set is that it is easier for analysts to model the conditional expectation
  correctly with a fewer variables --- the key necessary assumption for
  outcome-model-based estimators. We leave further technical and thorough
  investigation of outcome-model-based estimators for future work.}
and a doubly robust estimator \citep{hernan2019}. In particular, for
the outcome-model-based estimator, we use a fully-interacted linear
model. Within the experimental data, we estimate a linear regression
with a specified set of covariates separately for the treatment and
control groups. Then, we use the estimated models to predict potential
outcomes under treatment and control for the target population
data. This outcome-model-based estimator is consistent under the assumption
that the outcome model is correctly specified. For the doubly robust
estimator, we use an augmented IPW estimator \citep{robins1994,
  hernan2019} where the outcome model is a fully-interacted linear
model and the sampling model is a logistic regression specified in
Section~\ref{subsec:est_pate}. This doubly robust estimator is
consistent if one of the two models --- outcome or sampling models ---
is correctly specified.

We first extend our analyses in Section~\ref{subsec:exact-app}. Table~\ref{tab:tab-exact-out} reports
results based on the outcome-model-based estimator (an extension of
Table~\ref{tab:estimate_XH}). Similarly to the case of the IPW estimator,
we find that (1) point estimates based on estimated separating sets are
similar to those based on the original sampling set, and (2) standard
errors based on our proposed estimated separating sets are smaller for
16 out of 17 outcomes. Table~\ref{tab:tab-exact-aipw} reports
results based on the doubly robust estimator (an extension of
Table~\ref{tab:estimate_XH}). Similarly to the cases of the IPW estimator
and the outcome-model-based estimator, we find that (1) point estimates based on estimated separating sets are
similar to those based on the original sampling set, and (2) standard
errors based on our proposed estimated separating sets are smaller for
15 out of 17 outcomes. Therefore, for all three classes of estimators,
our proposed approach of using the estimated separating set improves
estimation accuracy. Finally, we also compare estimates across three
classes of estimators in Table~\ref{tab:tab-exact}. Across 17 outcomes, we find that
estimates of the PATE are relatively stable across different estimators (none of the differences
in estimates are statistically significant at the conventional $0.05$
level), which suggests model misspecification is of little
concern.

We next extend our analyses in Section~\ref{subsec:mar-app}.
Table~\ref{tab:tab-mar} reports results for
Section~\ref{subsec:mar-app} by comparing estimates from
the outcome-model-based estimator and the doubly robust estimator to estimates from the
IPW estimator. While the point estimate for ``Agricultural'' is
unstable due to a relatively large standard error (the first row in Table~\ref{tab:tab-mar}), estimates of the PATE are
relatively stable across different estimators (none of the differences
in estimates are statistically significant at the conventional $0.05$
level), which again suggests model misspecification is of little
concern.

\clearpage
\begin{table}[!h]
  \centering
  \scalebox{0.75}{
    \begin{tabular}{|l|cc|cc|}
      \hline
      & \multicolumn{2}{c|}{Original} & \multicolumn{2}{c|}{Estimated}  \\
      & \multicolumn{2}{c|}{Sampling Set} & \multicolumn{2}{c|}{Separating
                                            Set }  \\
      \hline
      & Estimate & S.E. & Estimate & S.E. \\
      \hline
      Average employment hours & 4.58 & 2.35 & 3.57 & 1.80 \\
      Agricultural & -0.00 & 1.61 & -1.22 & 1.45 \\
      Nonagricultural & 4.58 & 1.77 & 4.79 & 1.45 \\
      Skilled trades only & 3.70 & 1.03 & 4.08 & 0.86 \\
      No employment hours & -0.04 & 0.03 & -0.03 & 0.02 \\
      Any skilled trade & 0.27 & 0.05 & 0.25 & 0.04 \\
      Works mostly in a skilled trade & 0.02 & 0.02 & 0.04 & 0.02 \\
      Cash earnings & 5.20 & 7.31 & 8.22 & 7.02 \\
      Durable assets & 0.08 & 0.10 & 0.06 & 0.08 \\
      Vocational training & 0.52 & 0.05 & 0.50 & 0.04 \\
      Hours of vocational training & 250.32 & 34.71 & 280.24 & 27.43 \\
      Business assets & 340.79 & 141.74 & 367.61 & 127.23 \\
      Maintain records & 0.14 & 0.05 & 0.14 & 0.04 \\
      Registered & 0.03 & 0.04 & 0.04 & 0.03 \\
      Pays taxes & 0.01 & 0.05 & 0.02 & 0.05 \\
      Changed parish & 0.04 & 0.06 & -0.02 & 0.03 \\
      Lives in Urban area & -0.01 & 0.03 & -0.01 & 0.03 \\
      \hline
    \end{tabular}}
  \caption{Estimates of the PATEs based on Outcome-Model-Based
    Estimator, comparing the Original Sampling Set and Estimated Exact Separating
    Sets. Extension of Table~\ref{tab:estimate_XH} in Section~\ref{subsec:exact-app}.}\label{tab:tab-exact-out}
\end{table}

\begin{table}[!h]
 \centering
  \scalebox{0.75}{
    \begin{tabular}{|l|cc|cc|}
      \hline
      & \multicolumn{2}{c|}{Original} & \multicolumn{2}{c|}{Estimated}  \\
      & \multicolumn{2}{c|}{Sampling Set} & \multicolumn{2}{c|}{Separating
                                            Set }  \\
      \hline
      & Estimate & S.E. & Estimate & S.E. \\
  \hline
  Average employment hours & 2.49 & 3.16 & 2.48 & 2.73 \\
  Agricultural & -2.27 & 2.73 & -2.08 & 1.83 \\
  Nonagricultural & 5.10 & 2.99 & 4.56 & 2.24 \\
  Skilled trades only & 2.29 & 1.66 & 3.71 & 1.14 \\
  No employment hours & 0.01 & 0.03 & -0.01 & 0.03 \\
  Any skilled trade & 0.24 & 0.07 & 0.24 & 0.06 \\
  Works mostly in a skilled trade & -0.03 & 0.03 & 0.02 & 0.04 \\
  Cash earnings & 4.16 & 8.06 & 9.02 & 7.54 \\
  Durable assets & 0.02 & 0.15 & 0.14 & 0.15 \\
  Vocational training & 0.49 & 0.07 & 0.50 & 0.05 \\
  Hours of vocational training & 228.35 & 50.14 & 283.26 & 34.55 \\
  Business assets & 326.58 & 178.44 & 371.18 & 139.65 \\
  Maintain records & 0.16 & 0.07 & 0.17 & 0.07 \\
  Registered & 0.05 & 0.06 & 0.06 & 0.05 \\
  Pays taxes & -0.02 & 0.07 & 0.01 & 0.07 \\
  Changed parish & 0.06 & 0.07 & -0.04 & 0.05 \\
  Lives in Urban area & 0.01 & 0.05 & -0.01 & 0.04 \\
   \hline
    \end{tabular}}
   \caption{Estimates of the PATEs based on Doubly Robust
     Estimator, comparing the Original Sampling Set and Estimated Exact Separating
     Sets. Extension of Table~\ref{tab:estimate_XH} in Section~\ref{subsec:exact-app}.}\label{tab:tab-exact-aipw}
\end{table}

\begin{table}[!h]
  \centering
  \scalebox{0.875}{
  \begin{tabular}{|l|cc|cc|cc|}
    \hline
    & \multicolumn{2}{c|}{IPW} & \multicolumn{2}{c|}{Outcome-Model-based}  & \multicolumn{2}{c|}{AIPW}  \\
    & \multicolumn{2}{c|}{Estimator} & \multicolumn{2}{c|}{Estimator}  & \multicolumn{2}{c|}{Estimator} \\
    \hline
    & Estimate & S.E. & Estimate & S.E. & Estimate & S.E. \\
    \hline
    Average employment hours & 4.79 & 2.39 & 3.57 & 1.80 & 2.48 & 2.73 \\
    Agricultural & 0.30 & 1.69 & -1.22 & 1.45 & -2.08 & 1.83 \\
    Nonagricultural & 4.49 & 1.79 & 4.79 & 1.45 & 4.56 & 2.24 \\
    Skilled trades only & 4.36 & 0.99 & 4.08 & 0.86 & 3.71 & 1.14 \\
    No employment hours & -0.03 & 0.03 & -0.03 & 0.02 & -0.01 & 0.03 \\
    Any skilled trade & 0.27 & 0.06 & 0.25 & 0.04 & 0.24 & 0.06 \\
    Works mostly in a skilled trade & 0.04 & 0.03 & 0.04 & 0.02 & 0.02 & 0.04 \\
    Cash earnings & 12.54 & 5.11 & 8.22 & 7.02 & 9.02 & 7.54 \\
    Durable assets & 0.18 & 0.13 & 0.06 & 0.08 & 0.14 & 0.15 \\
    Vocational training & 0.53 & 0.05 & 0.50 & 0.04 & 0.50 & 0.05 \\
    Hours of vocational training & 337.59 & 40.77 & 280.24 & 27.43 & 283.26 & 34.55 \\
    Business assets & 425.02 & 135.65 & 367.61 & 127.23 & 371.18 & 139.65 \\
    Maintain records & 0.20 & 0.07 & 0.14 & 0.04 & 0.17 & 0.07 \\
    Registered & 0.09 & 0.05 & 0.04 & 0.03 & 0.06 & 0.05 \\
    Pays taxes & 0.05 & 0.05 & 0.02 & 0.05 & 0.01 & 0.07 \\
    Changed parish & -0.01 & 0.04 & -0.02 & 0.03 & -0.04 & 0.05 \\
    Lives in Urban area & -0.01 & 0.04 & -0.01 & 0.03 & -0.01 & 0.04 \\
    \hline
  \end{tabular}}
\caption{Estimates of the PATEs based on Estimated Exact Separating
  Sets for Three Estimators. Extension of
  Section~\ref{subsec:exact-app}.}\label{tab:tab-exact}
\end{table}

\begin{table}[!h]
\centering
  \begin{tabular}{|l|cc|cc|cc|}
    \hline
    & \multicolumn{2}{c|}{IPW} & \multicolumn{2}{c|}{Outcome-Model-based}  & \multicolumn{2}{c|}{AIPW}  \\
    & \multicolumn{2}{c|}{Estimator} & \multicolumn{2}{c|}{Estimator}  & \multicolumn{2}{c|}{Estimator} \\
    \hline
    & Estimate & S.E. & Estimate & S.E. & Estimate & S.E. \\
  \hline
    Agricultural & 0.64 & 1.63 & -1.10 & 1.31 & -1.32 & 1.56 \\
    Changed parish & 0.05 & 0.03 & 0.03 & 0.03 & 0.04 & 0.03 \\
    Lives in Urban area & -0.02 & 0.03 & -0.02 & 0.03 & -0.01 & 0.04 \\
   \hline
  \end{tabular}
  \caption{Estimates of the PATEs based on Estimated Marginal Separating
  Sets for Three Estimators. Extension of
  Section~\ref{subsec:mar-app}.}\label{tab:tab-mar}
\end{table}

\clearpage
\section{Simulation Studies}
\label{sec:sims}
We turn now to simulations to explore how well the proposed algorithm
can recover the PATE. We first verify that our proposed
algorithm can obtain a consistent estimator of the PATE. More importantly, we find that
estimators based on estimated separating sets often have similar
standard errors to the ones based on the true sampling set. Although
our approach introduces an additional estimation step of finding
separating sets to relax data requirements for the target population, it does not suffer from substantial efficiency loss. Both
results hold with and without user constraints on what variables can be measured in the target population.

\subsection{Simulation Design}
In this subsection, we articulate our simulation design step by step.
See the supplementary material for all the details on the simulation design.

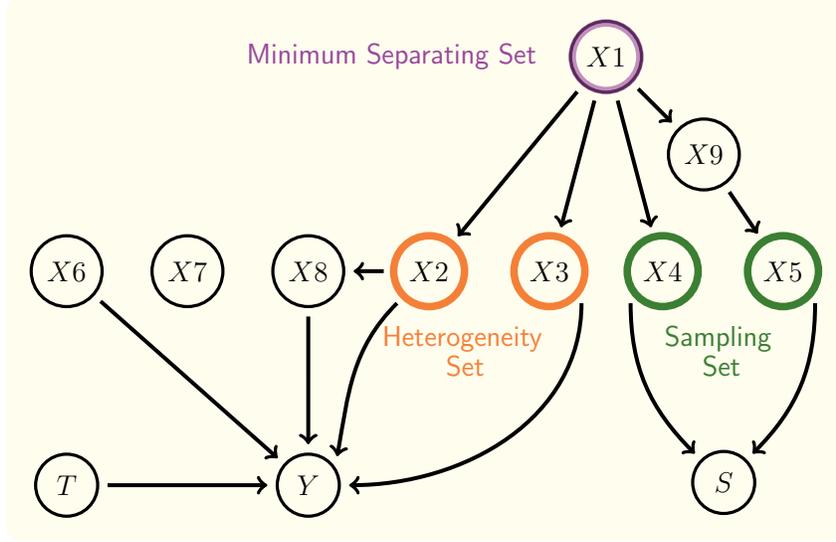
\begin{figure}[!t]
  \begin{center}
    \begin{tikzpicture}[roundnode/.style={circle, very thick, minimum size=11mm,inner sep=5,outer sep=5},
      roundnodecorr/.style={circle, line width = 1mm , minimum size=12mm,draw=Purple,opacity=0.6},
      myarrow/.style={->, line width = 1.5},
      mydoublearrow/.style={<->,style=dashed, line width = 1.5},
      font=\sffamily\Large,
      scale=0.75, every node/.style={scale=0.75}]
      \pgfplotsset{every axis/.append style={line width=1pt}}
      % nodes
      \node[roundnode, draw=black,text centered]   (X1)    {$X1$};
      \node[roundnodecorr] [right of= X1, node distance = 0mm]  	(X1corr)	{};
      \node[roundnode, draw=Orange, line width = 1mm, text centered] [below left = 20mm and 15mm of X1]  (X2)    {$X2$};
      \node[roundnode, draw=Orange, line width = 1mm, text centered] [below left = 20mm and -1mm of X1]  (X3)    {$X3$};
      \node[roundnode, draw=OliveGreen,  line width = 1mm, text centered] [below right = 20mm and -1mm of X1]  (X4)    {$X4$};
      \node[roundnode, draw=OliveGreen, line width = 1mm, text centered] [below right = 20mm and 15mm of X1]  (X5)    {$X5$};

      \node[roundnode,draw=black,text centered] [left = 4mm of X2]  (X8)    {$X8$};
      \node[roundnode,draw=black,text centered] [left = 4mm of X8]  (X7)    {$X7$};
      \node[roundnode,draw=black,text centered] [left = 4mm of X7]  (X6) {$X6$};
      \node[draw=none] [left of= X1, node distance = 38mm] (X1corr2)	{\textcolor{Purple}{Minimum Separating Set}};
      \node[draw=none] [below right = 2mm and -11.5mm of X2, node distance =
      38mm] (X2name)	{\textcolor{Orange}{Heterogeneity}};
      \node[draw=none] [below right = 6mm and -3mm of X2, node distance =
      38mm] (X22name)	{\textcolor{Orange}{Set}};
      \node[draw=none] [below right = 2mm and -5mm of X4, node distance = 38mm] (X4name)	{\textcolor{OliveGreen}{Sampling}};
      \node[draw=none] [below right = 6mm and 0mm of X4, node distance = 38mm] (X42name)	{\textcolor{OliveGreen}{Set}};

      \node[roundnode,draw=black,text centered] [above left = 7mm and 2mm of X5]  (X9)    {$X9$};

      \node[roundnode,draw=black,text centered] [below = 17mm of X8]  (Y)    {$Y$};
      \node[roundnode,draw=black,text centered] [below right = 20mm and 0mm of X4]  (S)    {$S$};

      \node[roundnode,draw=black,text centered] [below = 17mm of X6]  (T)    {$T$};

      \draw[myarrow,draw=black] (X1) -- (X2) {};
      \draw[myarrow,draw=black] (X1) -- (X3) {};
      \draw[myarrow,draw=black] (X1) -- (X4) {};

      \draw[myarrow,draw=black] (X1) -- (X9) {};
      \draw[myarrow,draw=black] (X9) -- (X5) {};

      \draw[myarrow, draw=black] (X4.south west)  to [out=270,in=135]  (S.north west) {};
      \draw[myarrow,draw=black] (X5.south east) to [out=270,in=45] (S.north east) {};

      % (X2) edge [bend left=10, draw=black] node[right=1mm]{} (Y);
      \draw[->, line width = 1.5, draw=black] (X2) to [out=225,in=80] (Y.north east) {};
      \draw[myarrow,draw=black] (X3.south east) to [out=270,in=0] (Y.east) {};

      \draw[myarrow,draw=black] (X6) -- (Y) {};

      \draw[myarrow,draw=black] (X8) -- (Y) {};
      \draw[myarrow,draw=black] (X2) -- (X8) {};

      \draw[myarrow,draw=black] (T) -- (Y) {};

      % \draw[myarrow,draw=Mahogany,dashed] (S) -- node [midway] {\Huge{\textcolor{Mahogany}{$\mathbf{\times}$}}} (Y);
      % \draw[myarrow,draw=Mahogany,dashed] (S) -- node [midway,below] {\textcolor{Mahogany}{Stability of Treatment}} (Y);

      % \node[draw=none] [above = 5mm of X7]  (xxx) {\textcolor{black}{Baseline}};
      % \draw[myarrow,draw=lightgray] (xxx) -- (X6) {};
      % \draw[myarrow,draw=lightgray] (xxx) -- (X8) {};

      \begin{pgfonlayer}{background}
        \filldraw [line width=4mm,join=round,yellow!8]
        (X1.north  -| X6.west)  rectangle (Y.south  -| X5.east);
      \end{pgfonlayer}

    \end{tikzpicture}
    \spacingset{1}{
      \caption{Causal DAG underlying the simulation
        study. \label{Dag_sep} Note: We consider three conceptually
        distinct sets (1) a sampling set, $X4$ and $X5$ (green),
        (2) a heterogeneity set, $X2$ and $X3$ (orange) and (3)
        the minimum separating set, $X1$ (purple).  Three root nodes $X1$, $X6$, $X7$ are normally distributed and other
        pre-treatment covariates are linear functions of their parents. }}
  \end{center}
\end{figure}

\paragraph{Pre-treatment Covariates and Potential Outcome Model.}
To consider different types of separating sets, we assume the causal
directed acyclic graph (DAG) in Figure \ref{Dag_sep} that encodes causal
relationships among the outcome, the sampling indicator, and
pre-treatment covariates. In this DAG, there are three conceptually
distinct sets that we consider -- (1) a sampling set, $X4$ and $X5$, depicted in green,
(2) a heterogeneity set, $X2$ and $X3$, depicted in orange, and (3)
the minimum separating set, $X1$, highlighted in purple. Three root nodes $X1$, $X6$, $X7$ are normally distributed and other
pre-treatment covariates are linear functions of their parents in the
DAG.  In particular, pre-treatment covariates are generated as follows.
\begin{align*}
  X1 &\sim \cN(0, 1) \\
  X2 & =  0.7 \times X1 + \sqrt{1 - 0.7^2} \times \epsilon_2 \\
  X3 & = 0.7 \times X1 + \sqrt{1 - 0.7^2} \times \epsilon_3 \\
  X4 & =  0.7 \times X1 + \sqrt{1 - 0.7^2} \times \epsilon_4 \\
  X5 & = 0.3 \times X9 + \sqrt{1 - 0.3^2} \times \epsilon_5 \\
  X6 & \sim \cN(0, 1) \\
  X7 &  \sim \cN(0, 1) \\
  X8 & =  -0.7 \times X2 + \sqrt{1 - 0.7^2} \times \epsilon_8 \\
  X9 & =  0.6 \times X1 + \sqrt{1 - 0.6^2} \times \epsilon_9
\end{align*}
where $\epsilon_2, \epsilon_3, \epsilon_4, \epsilon_5, \epsilon_8,
\epsilon_9$ are drawn independently and identically from a standard
normal distribution, $\cN(0, 1).$ This results in the following correlation structure for variables $X1 - X9$.
\begin{small}
  \begin{align*}
    cor(\textbf{X}) &=
                      \begin{pmatrix}
                        1.00 & -0.70  &  0.70  &  0.70  & -0.20  &  0.00 &   0.00 &   0.50  &  -0.70 \\
                        -0.70  &  1.00  & -0.50  & -0.50  &  0.15  &  0.00  &  0.00  & -0.70  &  0.50 \\
                        0.70  &  -0.50  &  1.00  &  0.50  & -0.15  &  0.00  &  0.00  &   0.33  & -0.50 \\
                        0.70  &  -0.50  &  0.50  &  1.00  & -0.15  &   0.00  &  0.00 &   0.33  &  -0.50 \\
                        -0.21  &  0.15  & -0.15  & -0.15   & 1.00  &  0.00   & 0.00  & -0.10  &  0.30 \\
                        0.00  &  0.00  &  0.00 &   0.00  &  0.00   & 1.00  &  0.00  &  0.00  &  0.00 \\
                        0.00   & 0.00  &  0.00  &  0.00  &  0.00  &  0.00   & 1.00   & 0.00  &  0.00 \\
                        0.50  &  -0.70  &  0.33  &  0.33  & -0.10  &  0.00  &  0.00  &  1.00  & -0.33 \\
                        -0.70   & 0.50  & -0.50  & -0.50   & 0.30  &  0.00  &  0.00  & -0.33   & 1.00
                      \end{pmatrix}
  \end{align*}
\end{small}
We then draw the potential outcomes as follows.
\begin{equation*}
  Y_i(T_i) = 5 T_i + 10 \times X_{3i} \times T_i -10  \times X_{2i} \times T_i  +  X_{6i} - 3 \times X_{8i} + \epsilon_i
\end{equation*}
where $\epsilon_i \sim N(0, 1)$. Thus, the true PATE is set to $5$.

\paragraph{Sampling Mechanism and Treatment Assignment.}
We randomly sample a set of $n$ units for a randomized experiment.
The sampling mechanism is a logit model based on the sampling set, $X4$
and $X5$. % with probability scaled to the corresponding experimental sample size
 The treatment assignment mechanism is defined only for
the experimental sample ($S_i=1$). After being sampled into the
experiment, every unit has the same probability of receiving the
treatment $\Pr(T_i = 1 \mid S_i=1) = 0.5$.  For the sake of
simplicity, we omit an arrow from the sampling indicator $S$ to the treatment $T$ in Figure 1.
% Details are available in Appendix \ref{app:simul}.

In particular, we draw a sampling indicator $S_i$ as follows.  The second step scales the probability to be bounded away from zero and one.
\begin{align*}
&  S'_{i,lp} = -20 \times X_{4i} + 20 \times X_{5i} \\
 & S_{i,lp} = 0.25(S'_{i,lp} - \overline{S'_{lp}})/sd(S'_{lp}) \\
 & S_i = \frac{1}{1 + e^{-S_{i,lp}}}
\end{align*}

\paragraph{Simulation Procedure.}  We conduct 5000 simulations for
each of six experimental sample sizes, $n = \{ 100, 200, 500, 1000,
2000, 3000 \}$. Within each simulation, we first randomly sample $n$
units for the experiment based on the sampling mechanism and randomly
assign units to treatment according to the specified treatment
assignment mechanism.  We also randomly sample a target population of
size $m = 10000$.  We then estimate both an exact and a marginal
separating set using the experimental data.  % Here, we use the causal
% DAG to clarify our simulation settings. In the estimation, we do not estimate this underlying causal DAG, but rather estimate the separating set using an MRF in order to find a set that meets the left-hand side criterion in Theorems \ref{thm_id_Exsep}  or \ref{thm_id_sep}.
% If multiple sets are discovered, select one at random.
An advantage of our method is that researchers can specify
variables that cannot be measured in the target population.  To
illustrate this benefit,  we also
estimate a marginal separating set with a constraint that variable
$X1$ is unmeasurable in the target population, thus making the minimal separating set unobservable in the target
population.  We compare these sets to an oracle sampling set, oracle
heterogeneity set, and oracle minimum separating set.

For each estimated and oracle set, we compute the PATE using the
inverse probability weighting estimator described in
Section~\ref{subsec:est_pate}.  In the supplementary material, we repeat these simulations with a calibration estimator discussed in
\citet{Hartman:2015hq}, and a linear regression projection estimator.

\begin{figure}
    \centering
    \includegraphics[width=0.9\textwidth]{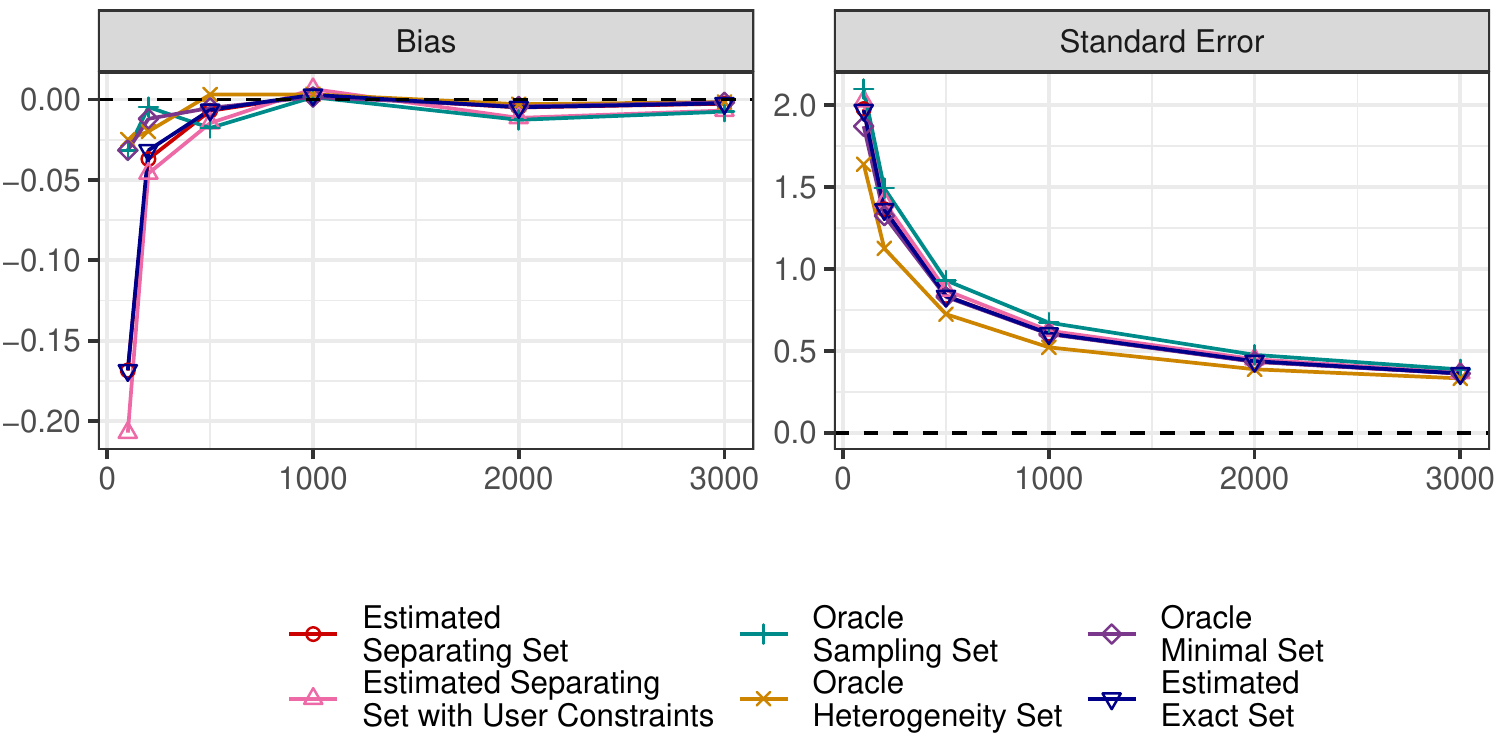}
    \spacingset{1}{\caption{Simulation Results.  Note: The left figure
        shows bias for the PATE and the right figure presents
        standard error estimates. As expected, bias is close to zero
        for all estimators. More importantly, estimators based on the estimated
        separating sets (red) and estimated separating sets with user constraints (pink)
        have similar standard errors to the oracle sampling
        set (green) and the oracle minimum separating set (purple).}
    \label{fig:sim_res_1}}
\end{figure}

\subsection{Results}
We present results in Figure~\ref{fig:sim_res_1}.  Not shown in the
graph are the results for the naive difference-in-means, which has
significant bias ($-1.0$).  As expected, we see that the bias goes
to zero for the oracle and estimated separating sets, and that the
estimators are consistent for the PATE.  More importantly, we see that
estimators based on the selected marginal separating sets (red), exact separating sets (dark blue), and marginal separating set with user constraints (pink) have similar standard errors to the oracle sampling
set (green) and the oracle minimum separating set (purple).  An estimator based
on the oracle heterogeneity set (orange) has smaller standard errors
than other estimators partly because it contains variables which are direct
predictors of outcomes. % However, this estimator might be unavailable
% in practice as discussed in Section~\ref{sec:dis} because a
% heterogeneity set is inherently unobservable.

\begin{figure}
    \centering
    \includegraphics[width=0.9\textwidth]{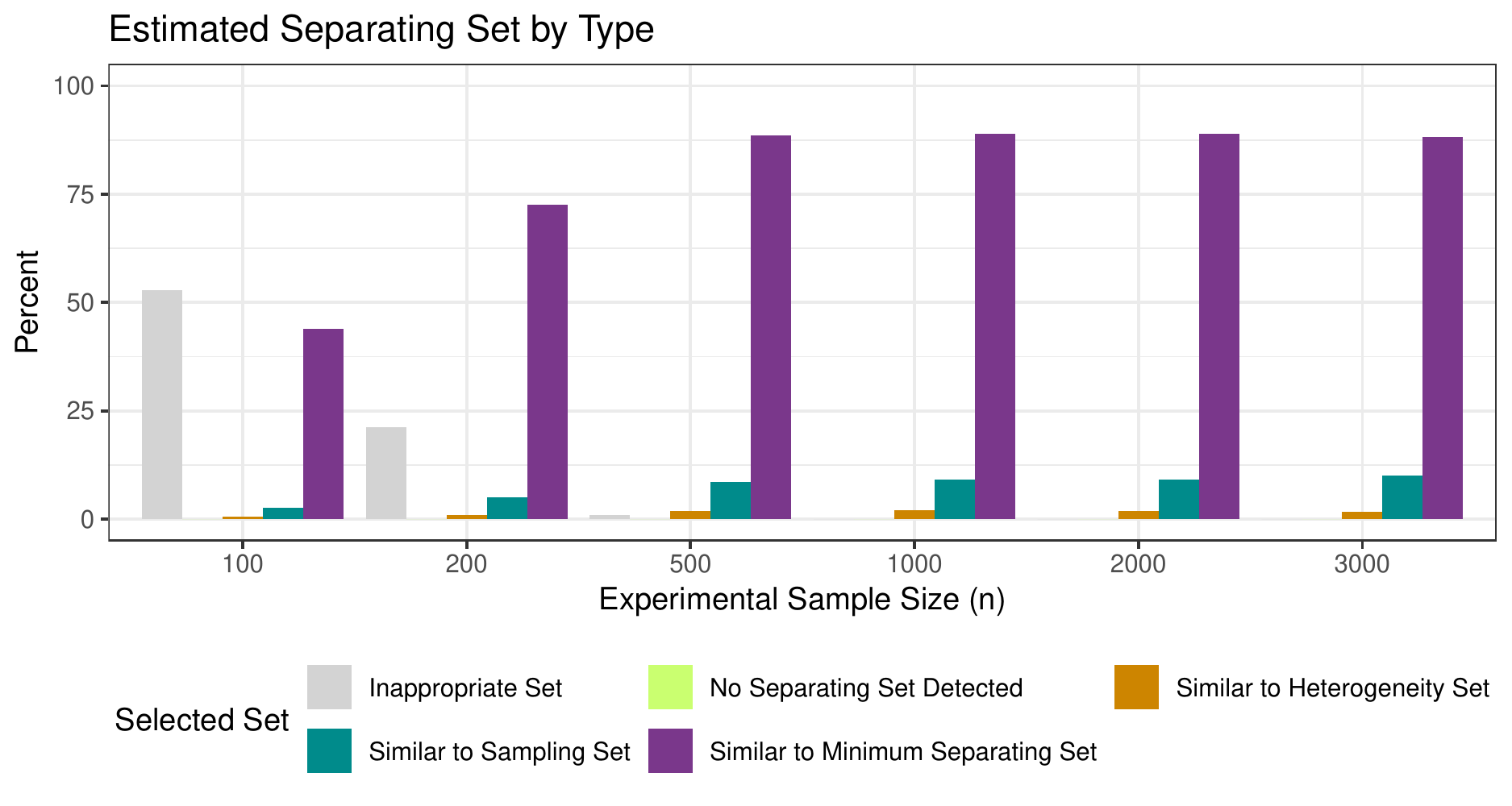}
    \spacingset{1}{\caption{Types of Estimated Separating Sets.  Note:
        We present the frequency of estimated separating sets by
        conceptual type.  While the algorithm picks an inappropriate set
        when the sample size is small, as $n$ increases, the most likely
        set is the minimal separating set.}\label{fig:sim_res_2_type}}
\end{figure}

Figure~\ref{fig:sim_res_2_type} shows the breakdown of types of
estimated separating sets. %\footnote{Appendix \ref{app:simulation_res} presents numerical results for bias and standard error by selected type of separating set.}
% Since the algorithm can discover multiple
% separating sets, we select one at random.
We group sets that are conceptually similar, and the frequency with which each set is chosen is
presented. For example,
if our algorithm selects the variables in the sampling set ($X4$ and
$X5$) as well as an additional variable, we group these as ``similar
to'' the sampling set.  As can be seen, in these simulations as $n$ gets large, over
75\% of the time, the minimal separating set (purple) is selected. % \footnote{If we do not include any thresholding to encourage sparsity in the MRF, the algorithm more frequently picks the sampling set.}
  Small sample size can lead to the misestimation
of the MRF, and therefore selection of inappropriate sets (gray) which do
not remove bias --- however, the rate at which inappropriate sets are
selected drops off rapidly with sample size.  In the supplementary
material, we show that, when incorporating user constraints that make adjustment by the minimum separating set infeasible, the algorithm selects sets
similar to the sampling and heterogeneity sets with higher frequency.

\subsection{Additional Simulation Results}
\label{app:simulation_res}
\begin{figure}
    \centering
    \includegraphics[width=\textwidth]{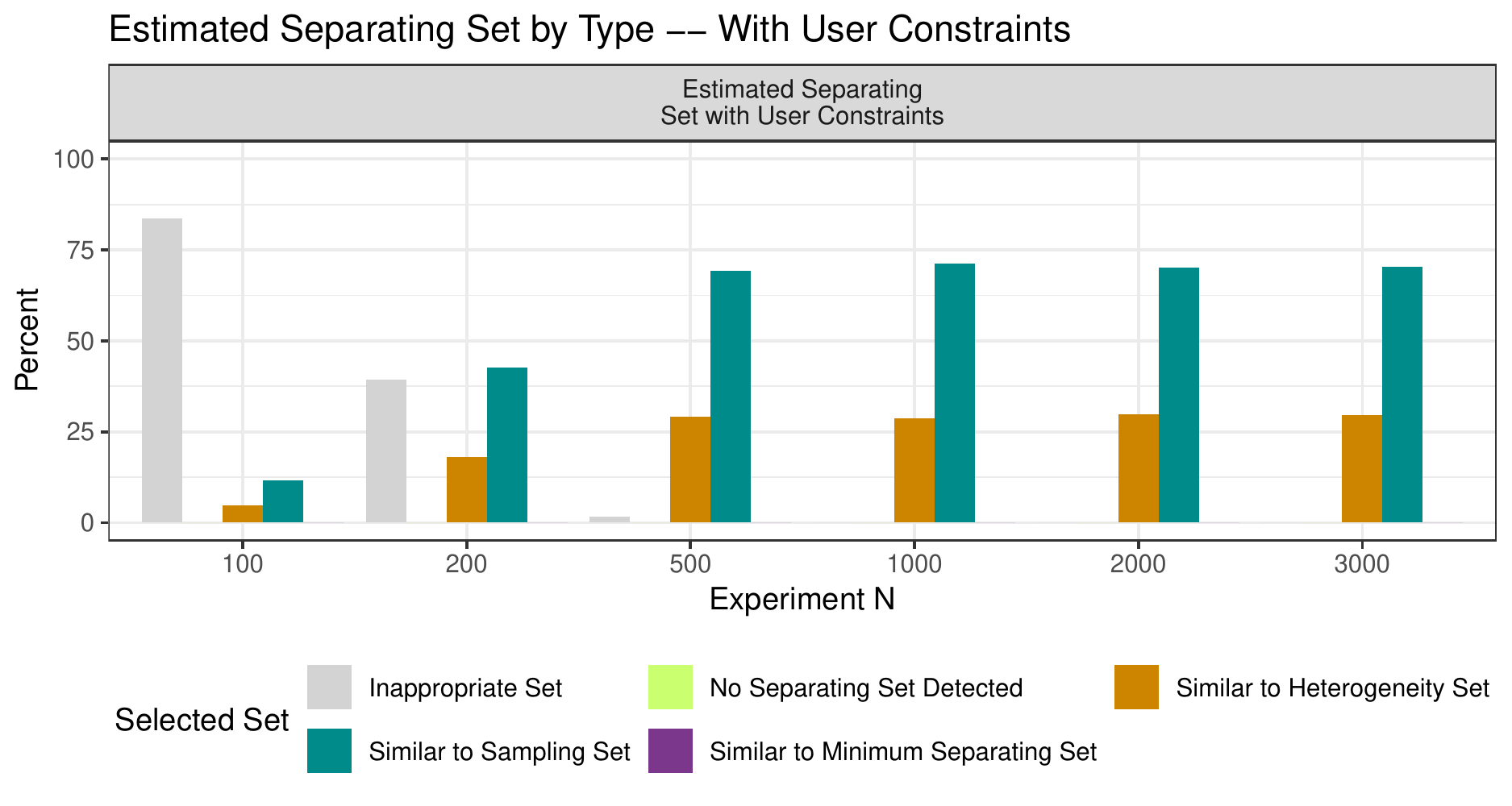}
    \spacingset{1}{\caption{Type of Estimated Marginal Separating Set with
        User Constraints.  Note: We present the frequency of estimated separating sets by
        conceptual type. With user constraints, the algorithm selects each of the other types of
        separating sets more frequently.}\label{fig:sim_res_2_type_unobs}}
\end{figure}

In the previous subsection, we discussed the breakdown of the different types of estimated separating sets
in the simulated data generating process.  Here we show the breakdown of types of estimated separating sets
when incorporating user constraints in Figure~\ref{fig:sim_res_2_type_unobs}.
In this case, $X1$, the alternative separating set, cannot be measured in the target population,
we see that the algorithm selects the sampling and heterogeneity sets with higher frequency.

% recreate the results from Figure~\ref{fig:sim_res_2_type}, comparing the results to when the sampling set is unmeasured in the population, in Figure~\ref{fig:sim_res_2_type_unobs}.  Once again, we present the frequency of which sets are estimated in the data, including when multiple sets are found.

\begin{figure}[!h]
    \centering
    \includegraphics[width=\textwidth]{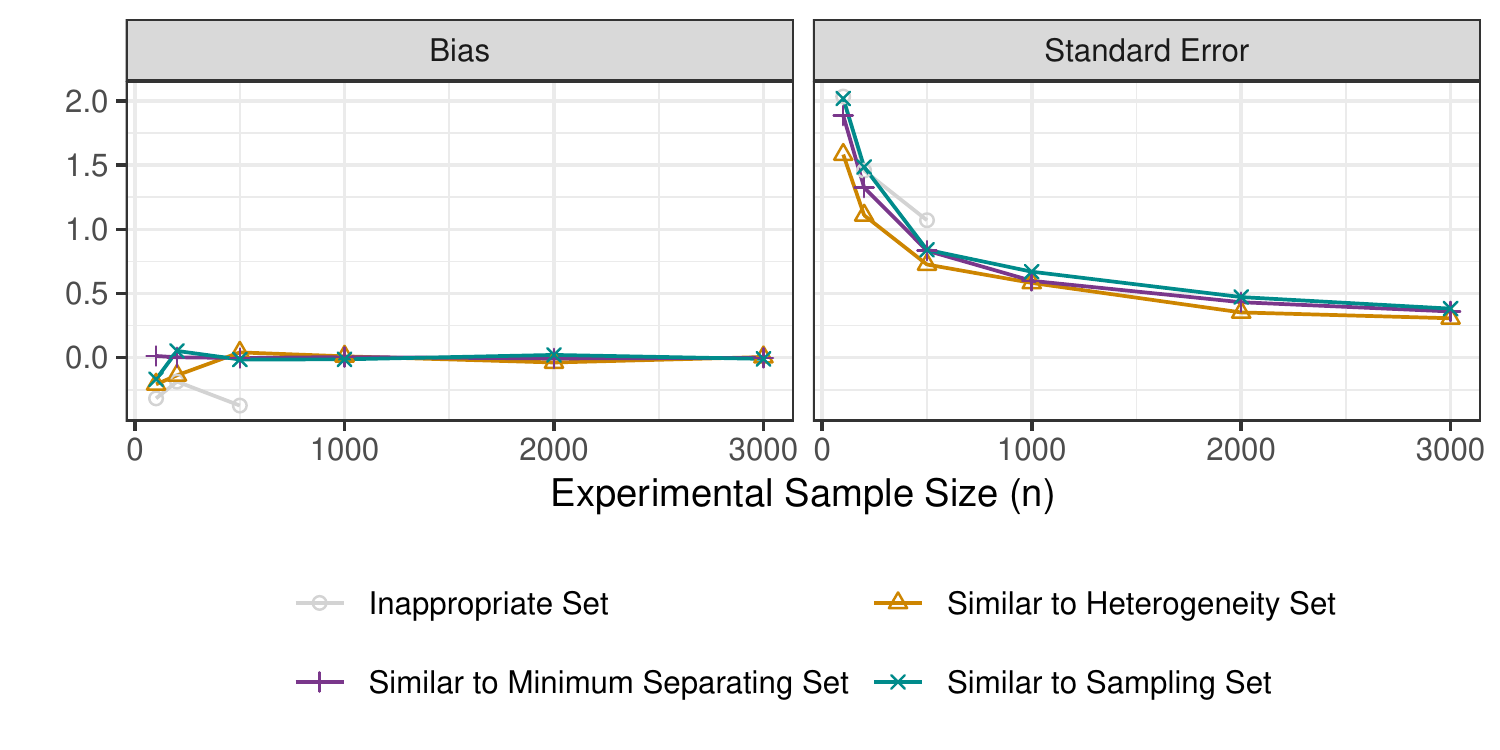}
    \spacingset{1}{\caption{Simulation Results for Estimated Separating Set by Type.  Note: The left figure
        shows bias for the PATE and the right figure presents
        standard error estimates.
        As expected, bias is close to zero
        for all estimators. Estimated sets are categorized by type: similar to oracle sampling
        set (green) and the oracle minimum separating set (purple) and oracle heterogeneity set (orange).}
    \label{fig:sim_res_estimated_by_type}}
\end{figure}

Figure \ref{fig:sim_res_estimated_by_type} presents the bias and standard error result by selected estimated
separating set type.  We refer to sets that are ``similar to''
different conceptual sets in order to group sets that control for a
specific type of separating sets, but which may include extra
variables.  For example, if the estimated set includes $X4$, $X5$, and
$X8$, we say this is similar to a sampling set ($X4$ and $X5$). As
theorems tell us, it doesn't matter what type of
separating sets the algorithm estimates in the experimental data, all
of them produce unbiased estimates so long as the set is an
appropriate separating set (see Figure~\ref{fig:sim_res_estimated_by_type}). When
an inappropriate set is chosen, which is common in the $n =
100$ case but rare as $n$ increases, we see that inappropriate sets do
not reduce bias. As we expect, when estimated separating sets are similar to a heterogeneity set, standard errors are the smallest.

\begin{figure}[!h]
    \centering
    \includegraphics[width=0.8\textwidth]{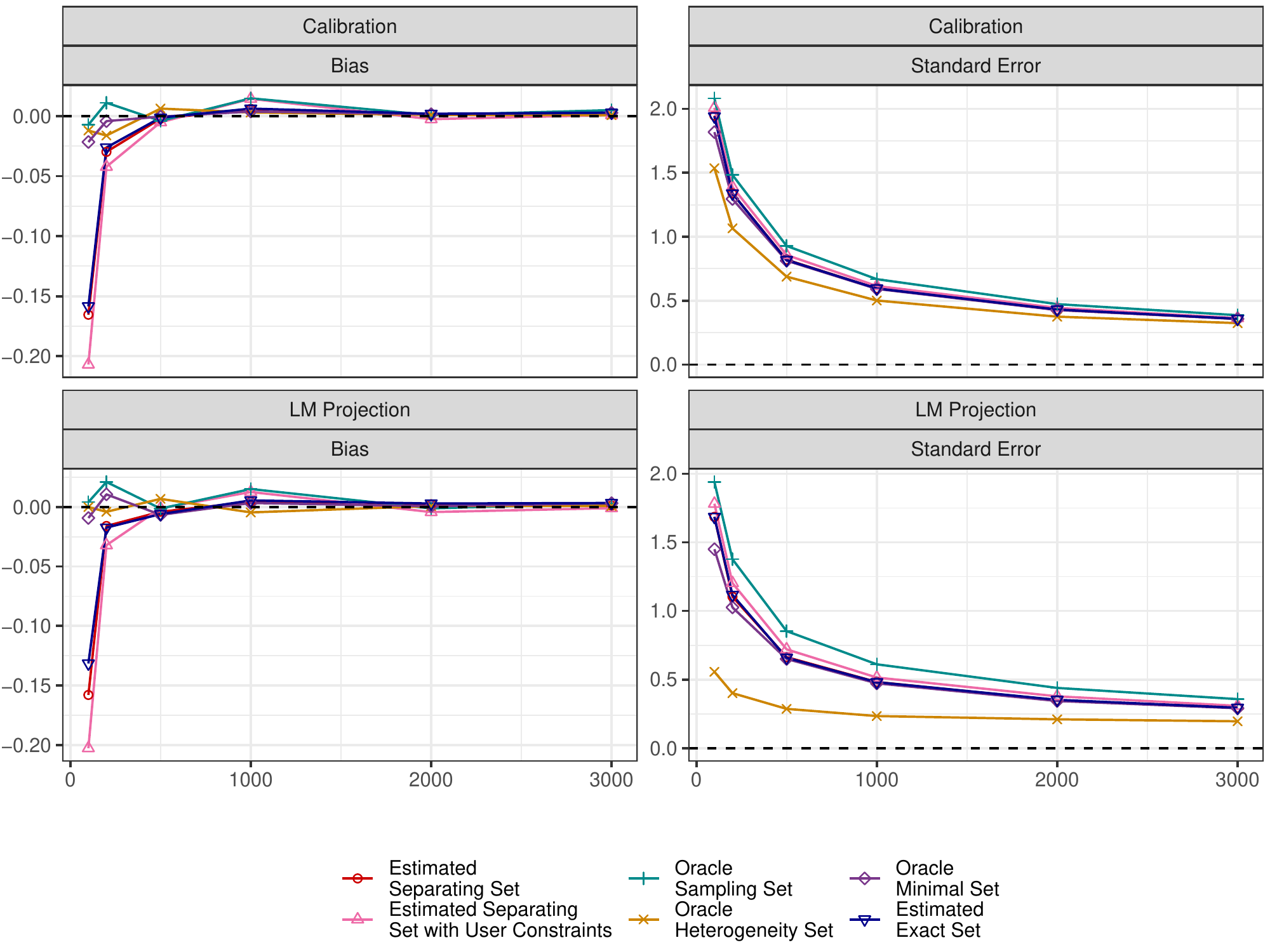}
    \spacingset{1}{\caption{Simulation Results for Alternative Estimators.  Note: The left figure
        shows bias for the PATE and the right figure presents
        standard error estimates.
        As expected, bias is close to zero
        for all estimators. More importantly, estimators based on the estimated
        separating sets (red) and estimated separating set with user constraints (pink)
        have similar standard errors to the oracle sampling
        set (green) and the oracle minimum separating set (purple). An estimator based
        on the heterogeneity set (orange) has significantly smaller standard errors
        than other estimators, but this estimator might be unavailable
        in practice.}
    \label{fig:sim_res_1_alt}}
\end{figure}

Finally, we present the simulation results for two alternative estimators in Figure \ref{fig:sim_res_1_alt}, a calibration estimator and a linear regression projection.
The calibration estimator matches population means for the estimated separating set using a maximum entropy (raking) algorithm \citep{Hartman:2015hq}.
The linear projection estimator estimates a fully interacted linear regression model using the estimated separating set, and projects the model on the
target population.

\clearpage
\section{R Function to Estimate Separating Sets}
{\footnotesize
\begin{verbatim}
# ################################
# Estimating the separating set
# ################################
# X.data: all pre-treatment covariates in the experimental data
# X.type: types of each covariate. "g" for continous variables, and "c" for categorical variables.
# X.level: the number of levels in each covariates. For continous variables, set it to 1.
# Y: outcome variable in the experimental data
# Treat: treatment variable in the experimental data
# XS: names of the sampling set
# XH: names of the heterogeneity set
# XU: names of variables unmeasurable in the target population
# type: when "Y", we estimate the marginal separating set. when "XH", we estimate the exact separating set.
# print_graph: whether we print the estimated Markov Random Fields

library(igraph); library(qgraph); library(lpSolve); library(mgm); library(Hmisc)

Separating <- function(X.data, X.type, X.level,
                       Y, Treat, XS, XH = NULL, XU=NULL, type = "Y",
                       print_graph = FALSE) {

  ## Setup
  n.var <- ncol(X.data)
  if(type == "Y"){
    if(missing(X.type) == TRUE){
      type.sim  <- rep("g", n.var + 2)
      level.sim <- rep(1, n.var + 2)
    }else{
      type.sim <- c(X.type, rep("g", 2))
      level.sim <- c(X.level, rep(1, 2))
    }
    X.data.g <- cbind(X.data, Y, Treat)
    name.label <- c(colnames(X.data), "Y")
  }else if(type == "XH"){
    if(missing(X.type) == TRUE){
      type.sim  <- rep("g", n.var + 1)
      level.sim <- rep(1, n.var + 1)
    }else{
      type.sim <- c(X.type, rep("g", 1))
      level.sim <- c(X.level, rep(1, 1))
    }
    X.data.g <- cbind(X.data, Treat)
    name.label <- colnames(X.data)
  }



  ## ###########################################
  ## Step 1: Estimate the Markov Random Graph
  ## ###########################################
  fit.sim <- mgm(
    data = X.data.g,
    type = type.sim,
    level = level.sim,
    threshold = "LW",
    k = 2,
    verbatim = TRUE,
    signInfo = FALSE,
    lambdaSel = "EBIC"
  )


  ## Remove T from the Graph
  treat_ind <- which(colnames(X.data.g) == "Treat")
  Ad <- as.matrix(fit.sim$pairwise$wadj > 0)
  Ad <- Ad[-treat_ind,-treat_ind]
  Ad.w <- fit.sim$pairwise$wadj
  Ad.w <- Ad.w[-treat_ind, -treat_ind]
  edge.col <- fit.sim$pairwise$edgecolor
  edge.col <- edge.col[-treat_ind,-treat_ind]
  graph.u <- graph_from_adjacency_matrix(Ad)

  ## Show the graph
  if(print_graph) qgraph(
    Ad.w,
    edge.color = edge.col,
    layout = 'spring',
    labels = name.label
  )

  ## #################################################################
  ## Step 2: Estimate the Separating Set based on an estimated MRF
  ## #################################################################
  if (type == "Y") {
    base <- rep(0, (n.var + 1))
    XS.ind <- which(is.element(colnames(X.data), XS))
    path.cons <- matrix(NA, nrow = 0, ncol = (n.var + 1))
    ## Enumerate all path
    for (w in 1:length(XS)) {
      ind.path.mat <- do.call("rbind",
                              lapply(all_simple_paths(graph.u, (n.var + 1), XS.ind[w]),
                                     FUN=function(x) ind.path(x, base)))
      path.cons <- rbind(path.cons, ind.path.mat)
    }
  }else if (type == "XH") {
    base <- rep(0, n.var)
    XJ  <- intersect(XS, XH)
    all.pair <- expand.grid(XH, XS)
    path.cons <- matrix(NA, nrow = 0, ncol = n.var)
    ## Enumerate all path
    for (w in 1:nrow(all.pair)) {
      ind_1 <- which(colnames(X.data.g) == all.pair[w, 1])
      ind_2 <- which(colnames(X.data.g) == all.pair[w, 2])
      ind.path.mat <-
        do.call("rbind",
                lapply(
                  all_simple_paths(graph.u, ind_1, ind_2),
                  FUN=function(x) ind.path(x, base)))
      path.cons <- rbind(path.cons, ind.path.mat)
    }
  }

  if(dim(path.cons)[1] == 0) {
    solution <- NULL
    status <- 0
  }else{

    ## Removing Y and XU from the separating set
    if (length(XU) == 0) {
      if(type == "Y"){
        path.cons2 <- rbind(path.cons,
                            c(rep(0, n.var), 1))
        f.dir <- c(rep(">=", nrow(path.cons)), "=")
        f.rhs <- c(rep(1, nrow(path.cons)), 0)
      }else if(type == "XH"){
        path.cons2 <- path.cons
        f.dir <- rep(">=", nrow(path.cons))
        f.rhs <- rep(1, nrow(path.cons))
      }
    } else{
      XU.ind <- which(is.element(colnames(X.data), XU))
      path.cons2.u <- matrix(0, nrow = length(XU.ind), ncol = n.var)
      for (i in 1:nrow(path.cons2.u)) {
        path.cons2.u[i, XU.ind[i]] <- 1
      }
      if(type == "Y"){
        path.cons2.u2 <- cbind(path.cons2.u, 0)
        path.cons2 <- rbind(path.cons,
                            c(rep(0, n.var), 1),
                            path.cons2.u2)
        f.dir <- c(rep(">=", nrow(path.cons)),
                   rep("=", (nrow(path.cons2.u2) + 1)))
        f.rhs <- c(rep(1, nrow(path.cons)),
                   rep(0, (nrow(path.cons2.u2) + 1)))
      }else if(type == "XH"){
        path.cons2.u2 <- path.cons2.u
        path.cons2 <- rbind(path.cons,
                            path.cons2.u2)
        f.dir <- c(rep(">=", nrow(path.cons)),
                   rep("=", nrow(path.cons2.u2)))
        f.rhs <- c(rep(1, nrow(path.cons)),
                   rep(0, nrow(path.cons2.u2)))
      }
    }

    if(type == "Y"){f.obj <- c(rep(1, n.var), 0)}
    else if(type == "XH"){f.obj <- rep(1, n.var)}
    f.con <- path.cons2

    num.solutions <- max.solutions.calculate <- 1
    sp.out <- lp("min", f.obj, f.con, f.dir, f.rhs, all.bin = TRUE, num.bin.solns = max.solutions.calculate)
    if(sp.out$status == 0) {
      if(max.solutions.calculate > 1) {
        solution <- sp.out$solution[1:(length(f.obj)*max.solutions.calculate)]
        solution <- split(solution, sort(1:length(solution) %% sp.out$num.bin.solns))
        if(num.solutions == 1) {
          solution <- sample(solution, num.solutions)
        }
        if(length(solution) == 1) {
          solution <- as.vector(unlist(solution))
        }
      } else {
        solution <- sp.out$solution
      }
    }
    status <- sp.out$status
  }

  ## Final Adjustment
  if (status == 0) {
    if(is.null(solution)==TRUE) {
      ## the empty set is enough for generalizability
      solution.name <- NULL
    }else{
      if(type == "Y"){
        solution.ind <-  which(solution[-length(solution)] == 1)
        XJ <- NULL
      }else if (type == "XH") {
        solution.ind <-  which(solution == 1)
      }
      solution.name <- colnames(X.data)[solution.ind]
      if(type == "XH" & length(XJ)!=0){ solution.name <- union(solution.name, XJ)}
    }
  }else if (status==2){
    cat("\nNo Feasible Solution.\n")
    solution.name <- "No Feasible Solution."
  }

  if(print_graph==TRUE){cat ("\n"); cat(solution.name)}

  return(solution.name)
}

# Auxiliary function
ind.path <- function(x, base) {
  base[x] <- 1
  return(base)
}
\end{verbatim}
}

\clearpage
\spacingset{1.4}
{\small
\pdfbookmark[1]{Supplementary Material References}{Supplementary Material References}
\printbibliography[title = Supplementary Material References]}

\end{refsection}

\end{document}